\documentclass[11pt, a4paper]{article}

\usepackage[T2A,T1]{fontenc}  
\usepackage[utf8]{inputenc}
\usepackage{lmodern}          
\usepackage{textcomp}
\usepackage[british]{babel}

\usepackage[a4paper, margin=25mm]{geometry}
\usepackage{microtype}

\usepackage{amsmath}
\usepackage{amssymb}          
\usepackage{siunitx}
\usepackage{graphicx}
\usepackage{xcolor}
\graphicspath{{figures/}}
\usepackage{tikz}
\usetikzlibrary{positioning, arrows.meta, fit, calc, shapes.geometric, decorations.pathreplacing}
\usepackage{caption}
\usepackage{booktabs}
\usepackage{array}
\usepackage{longtable}
\usepackage{tabularx}
\usepackage{rotating}
\usepackage{rotfloat}
\newcolumntype{C}{>{\centering\arraybackslash}X}
\newcolumntype{R}{>{\raggedleft\arraybackslash}X}
\newcolumntype{L}{>{\raggedright\arraybackslash}X}

\usepackage{enumitem}
\setlist{itemsep=2pt, topsep=4pt}
\usepackage{csquotes}        
\MakeOuterQuote{"}

\usepackage[
  backend=bibtex,
  style=numeric-comp,
  sorting=none,
  maxnames=6,
  minnames=1,
  giveninits=true,
  doi=true,
  url=false,
  isbn=false,
  eprint=false,
]{biblatex}
\usepackage[colorlinks=true,
            linkcolor=black!70!blue,
            citecolor=black!70!blue,
            urlcolor=black!70!blue]{hyperref}
\usepackage[capitalise, noabbrev]{cleveref}

\usepackage{newunicodechar}
\newunicodechar{⁻}{\textsuperscript{\textminus}}
\newunicodechar{⁰}{\textsuperscript{0}}
\newunicodechar{¹}{\textsuperscript{1}}
\newunicodechar{²}{\textsuperscript{2}}
\newunicodechar{³}{\textsuperscript{3}}
\newunicodechar{⁴}{\textsuperscript{4}}
\newunicodechar{⁵}{\textsuperscript{5}}
\newunicodechar{⁶}{\textsuperscript{6}}
\newunicodechar{⁷}{\textsuperscript{7}}
\newunicodechar{⁸}{\textsuperscript{8}}
\newunicodechar{⁹}{\textsuperscript{9}}
\newunicodechar{₀}{\textsubscript{0}}
\newunicodechar{₁}{\textsubscript{1}}
\newunicodechar{₂}{\textsubscript{2}}
\newunicodechar{₃}{\textsubscript{3}}
\newunicodechar{₄}{\textsubscript{4}}
\newunicodechar{₅}{\textsubscript{5}}
\newunicodechar{₆}{\textsubscript{6}}
\newunicodechar{₇}{\textsubscript{7}}
\newunicodechar{₈}{\textsubscript{8}}
\newunicodechar{₉}{\textsubscript{9}}

\usepackage{xspace}
\providecommand{\etal}{\textit{et~al.}\xspace}

\providecommand{\degC}{\si{\degreeCelsius}}
\providecommand{\um}{\si{\micro\metre}}
\providecommand{\nm}{\si{\nano\metre}}
\providecommand{\fs}{\si{\femto\second}}
\providecommand{\ps}{\si{\pico\second}}
\providecommand{\ns}{\si{\nano\second}}
\providecommand{\W}{\si{\watt}}
\providecommand{\kW}{\si{\kilo\watt}}
\providecommand{\Hz}{\si{\hertz}}
\providecommand{\kHz}{\si{\kilo\hertz}}
\providecommand{\MHz}{\si{\mega\hertz}}
\providecommand{\GHz}{\si{\giga\hertz}}
\providecommand{\THz}{\si{\tera\hertz}}

\title{Programmable vs. Static Beam Shaping in\\Ultrafast Laser Micromachining: A Critical Review}

\author{
  K.~Kobliha\textsuperscript{1,2,}\thanks{\,Corresponding author: krystof.kobliha@hilase.cz. ORCID: 0009-0002-1389-3234.}
  \and
  P.~Hauschwitz\textsuperscript{1,}\thanks{\,ORCID: 0000-0001-6243-0450.}
}

\date{
  \textsuperscript{1}~HiLASE Centre, Institute of Physics,
  Academy of Sciences of the Czech Republic, Za Radnici 828, 252 41 Dolni Brezany,
  Czech Republic\\
  \textsuperscript{2}~Faculty of Nuclear Sciences and Physical Engineering,
  Czech Technical University in Prague, Brehova 7, 115 19 Prague, Czech Republic
}

\begin{document}
\maketitle

\begin{abstract}
Beam shaping has become one of the principal determinants of throughput, precision, and process robustness in ultrafast laser micromachining. Despite this, the field is still largely interpreted through a historical distinction between programmable and static optical elements, a framework that increasingly fails to explain recent advances. This review re-examines that perspective and argues that beam shaping should instead be understood as a hardware–algorithm co-design problem. Across high-power spatial light modulators, machine-learning holography, hybrid optical architectures, and massively parallel processing, recent advances converge on the same conclusion: performance depends more on the co-design of optical hardware and computational algorithms than on any individual optical component. To establish a common basis for comparison, seven beam-shaping technologies and five algorithm families are evaluated within a unified seven-axis framework spanning optical performance, programmability, computational cost, and industrial deployment. This analysis identifies where the long-standing trade-off between throughput and flexibility has genuinely disappeared. In industrial parallel ablation and high-throughput two-photon polymerisation, programmable devices now sustain average powers once reserved for static optics, demonstrating why hardware-centred comparisons no longer capture the state of the art. Beyond reviewing recent developments, this work provides a predictive design framework for the next generation of beam-shaping systems. It introduces a benchmarking methodology, practical technology-selection criteria, measurable research milestones, and a~reporting standard for improving comparability across future studies.
\end{abstract}

\vspace{0.6em}
\noindent\textbf{Keywords:} beam shaping; spatial light modulator; laser micromachining; holography; two-photon polymerisation; diffractive optics; deep learning.

\vspace{0.6em}
\noindent\textbf{Highlights}
\begin{itemize}\setlength\itemsep{0pt}
  \item Programmable and static beam shaping are compared on the same seven axes.
  \item Performance is driven by designing the modulator and algorithm as a pair.
  \item The throughput-flexibility trade-off has collapsed only on the average-power axis.
  \item No programmable device yet has a characterised peak-fluence limit.
  \item Six testable milestones and a common reporting standard are set out.
\end{itemize} \newpage

\section{Introduction}
\label{sec:intro}

Beam shaping has become one of the principal determinants of throughput, precision, and process robustness in ultrafast laser micromachining. For more than two decades, however, it has been interpreted as a competition between programmable and static optical elements. Programmable shapers offer flexibility but limited power handling, while static optics deliver industrial throughput at the cost of reconfigurability \cite{xuLightFieldModulation2023, kazanskiyExploringFunctionalCharacteristics2024}. Historically, this distinction reflected genuine hardware limitations. Programmable devices lacked the power handling and update speed required for industrial processing \cite{buskeAdvancedBeamShaping2022, zhangFundamentalsPhaseonlyLiquid2014}, whereas static optics could not be reconfigured once fabricated \cite{fengSimplifiedFreeformOptics2017, tillkornAnamorphicBeamShaping2018}. This distinction became the dominant framework of the field, shaping both the development of beam-shaping technologies and the way they were compared \cite{xuLightFieldModulation2023, kazanskiyExploringFunctionalCharacteristics2024}. This review argues that this framework is no longer sufficient. The decisive factor is increasingly the co-design of the optical hardware and the algorithm that drives it, rather than the optical element itself. Performance is increasingly determined by the hardware–algorithm system rather than by any individual component.

This shift reflects broader advances in ultrafast laser processing. As laser sources have reached industrial average powers and megahertz repetition rates, the performance bottleneck has moved from the source to the shape of the energy delivered at the workpiece. Ultrafast laser micromachining shapes glass, dielectrics, semiconductors, and metals, often below the diffraction limit \cite{raciukaitisUltraShortPulseLasers2021, jiaRecentProgressFemtosecond2023}. Femtosecond and picosecond pulses deposit energy nonlinearly, so the heat-affected zone stays narrow, and the process barely depends on the substrate's linear absorption \cite{shinReviewHighprecisionFemtosecond2024}. The technique is now used across a wide range of industries. Examples include glass cutting and stealth dicing in electronics \cite{kimStudyGlassTGV2023, dudutisIndepthComparisonConventional2020, chengFlexibleTunedMultifocus2024}, surface texturing for wettability and tribology \cite{koblihaSuperhydrophobicSelfcleaningAluminium2025, hauschwitzRapidLaserinducedNanostructuring2025, yongReviewFemtosecondLaserInducedUnderwater2018}, buried-waveguide writing \cite{heSlitBeamShaping2022, liuFabricationSinglemodeCircular2021}, and two-photon polymerisation (TPP) of three-dimensional structures \cite{zhangHighThroughputTwoPhoton3D2024, balenaRecentAdvancesHighSpeed2023, zylaFrontiersLaserBased3D2024}.

A galvo-scanned Gaussian beam is often not the ideal profile for these applications. It~concentrates fluence at the centre and its wings extend well beyond the target contour, which in ablation produces overcut, sidewall taper, and an enlarged heat-affected zone compared with a flat-top or a task-matched profile \cite{hafnerTailoredLaserBeam2018, schmidtDynamicBeamShaping2024}. Such profiles can be generated using two broad classes of optical element. A static shaper, such as a $\pi$-shaper, axicon, or freeform element, imposes the wanted profile at high efficiency and full industrial power, but its design cannot be changed once made \cite{fengSimplifiedFreeformOptics2017, tillkornAnamorphicBeamShaping2018}. A programmable device, such as a liquid-crystal-on-silicon spatial light modulator (LCoS-SLM), can display an arbitrary phase pattern and update it on demand, but has historically done so only at a few tens of watts and at update rates far below an industrial pulse train \cite{buskeAdvancedBeamShaping2022, lutzEfficientUltrashortPulsed2021}.

\begin{figure}[!htbp]
  \centering
  \includegraphics[width=0.98\textwidth]{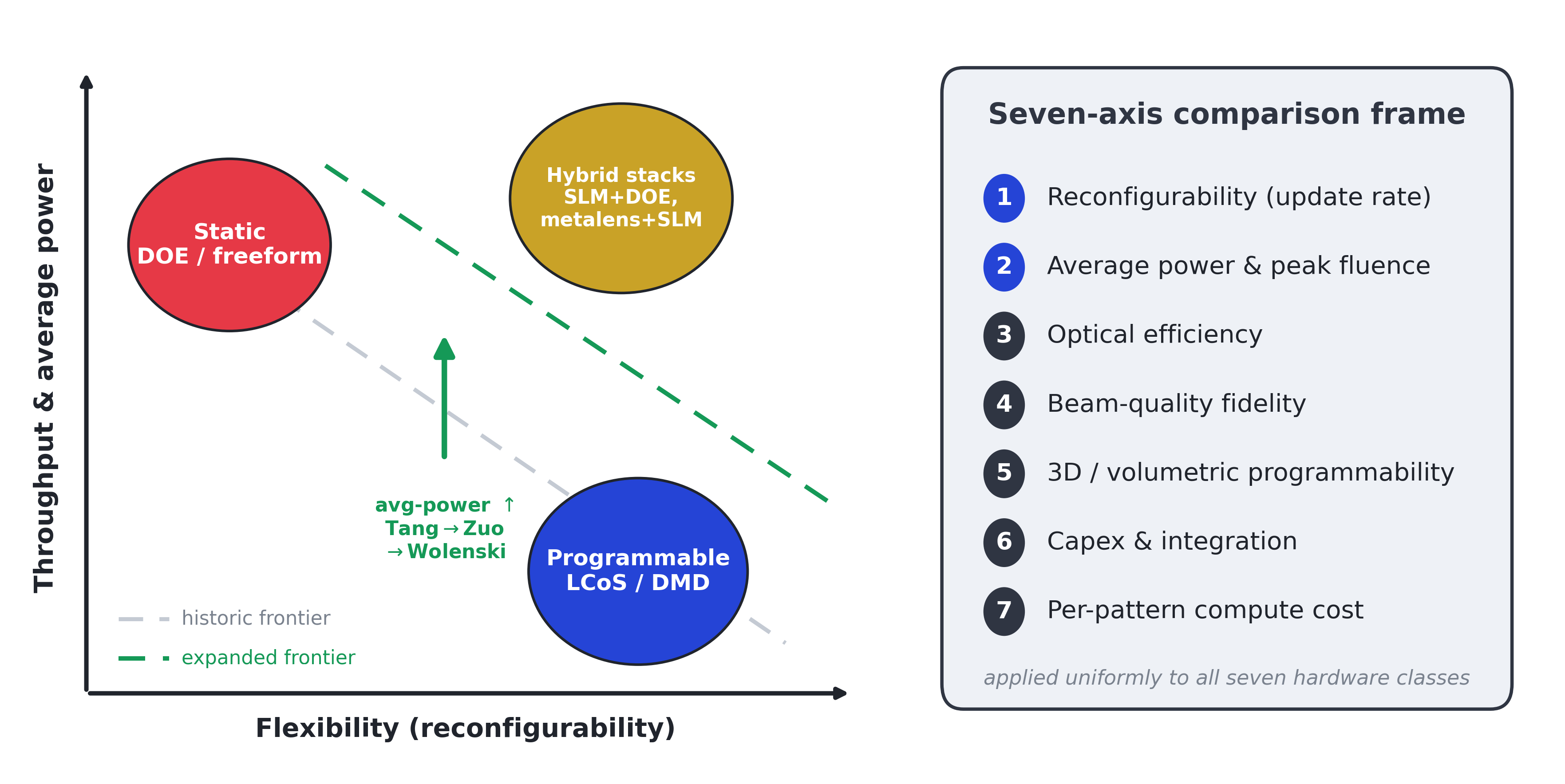}
  \caption{The throughput--flexibility plane. \emph{Left:} static optics (diffractive optical elements, freeform) occupy the high-throughput, low-flexibility corner and programmable modulators (LCoS-SLM, digital micromirror devices) the high-flexibility, historically low-throughput corner, separated by the historic frontier (grey dashed). Three developments of the last three years push it outward (green dashed): higher average power (section~\ref{sec:hardware}), fast machine-learning holography (section~\ref{sec:algorithms}), and hybrid stacks (section~\ref{sec:hybrid}). \emph{Right:} the seven-axis frame of section~\ref{sec:methods}, applied to every hardware class.}
  \label{fig:frontier_roadmap}
\end{figure}

The two families map onto a single frontier between throughput and flexibility, sketched in figure~\ref{fig:frontier_roadmap}. Static shapers sit in the high-throughput, high-power, inflexible corner; programmable modulators sit in the flexible, low-throughput corner. A decade ago the programmable envelope reached roughly $10^{2}$~\W\ of average power and update rates below the kilohertz, while static optics handled multi-kilowatt loads but could not be reprogrammed at all \cite{tillkornAnamorphicBeamShaping2018}. The frontier has a second, often overlooked dimension: even when hardware can switch patterns rapidly, generating the corresponding phase pattern has historically remained computationally expensive.

Three developments of the last three years now push this frontier outward. First, programmable hardware has moved up in average power. Tang \etal \cite{tangExtendingOperationalLimit2025} recovered a full $2\pi$ phase range up to 210~\W\ on a cooled SLM, Zuo \etal \cite{zuoHighPerformanceNIRLaserBeam2025} reached 0.98 first-order efficiency at 300~\W, and Wolenski \etal \cite{wolenskiKilowattAveragePower2025} reported continuous-wave (CW) operation at 1.4~\kW. Second, the algorithms have moved from iterative phase retrieval to learned inference. Where the Gerchberg--Saxton family and its mixed-region refinements iterate, a trained network outputs a pattern in one forward pass \cite{yuUseDeepLearning2025, eybposhDeepCGH3DComputergenerated2020, buskeAdvancedBeamShaping2022}, now demonstrated at 24.89 frames per second (FPS) on a $3840\times2160$ modulator \cite{yeomHighqualityPhaseonlyFourier2025}. 

\newpage 

Third, hybrid architectures combine programmable and static optical elements so that each compensates for the other's limitations. Gu \etal \cite{gu3DNanolithographyMetalens2025} illuminated a centimetre-scale metalens array with an upstream SLM to write TPP structures with more than $120{,}000$ foci. Kiefer \etal \cite{kieferMultiphoton77focus2024} reached an order $10^{8}$~voxels~s$^{-1}$ with a cascade of a diffractive element and a microlens array. Jacob \etal \cite{jacobDynamicBeamShaping2025} demonstrated a diffractive network that adapts across wavelengths, and reconfigurable metasurfaces have begun to appear \cite{crottiGiantUltrafastDichroism2024, hanFemtosecondLaserNonDiffractingBeam2026, wangTracingFootprintsFemtosecond2026}. None of these developments is captured by the traditional distinction between active and passive optics.

These developments share a common feature: progress increasingly depends on the co-design of optical hardware and computational algorithms rather than on advances in either alone. A~higher-power modulator is valuable only when an algorithm can generate a matching phase pattern, while learned holography is typically trained for a specific device \cite{pengNeuralHolographyCameraintheloop2020}. Likewise, hybrid architectures optimise programmable and static elements together \cite{liaoDifferentiableDesignFreeform2024}. Accordingly, hardware (section~\ref{sec:hardware}) and algorithms (section~\ref{sec:algorithms}) are treated here as a single design space rather than as separate topics.

Accordingly, the review is built around one question. Has the frontier between throughput and flexibility, which has defined beam shaping for two decades, begun to break down? To answer it, spatial light modulators and the algorithms that program them are set against diffractive optical elements, metasurfaces, deformable mirrors and freeform optics. Every hardware class, algorithm family and application is assessed against the same seven axes defined in section~\ref{sec:methods}: reconfigurability, average-power and peak-fluence handling, optical efficiency, beam-quality fidelity, three-dimensional programmability, capital cost, and per-pattern computational cost. The first six recur in earlier comparative reviews \cite{khoninaPerspectiveArtificialIntelligences2024, xuLightFieldModulation2023, kazanskiyExploringFunctionalCharacteristics2024}; the seventh is added here. Recent surveys each cover only one face of the problem: dynamic beam shaping with the liquid-crystal SLM \cite{mauclairDynamicSpatialBeam2025}, the light-field modulation algorithms \cite{xuLightFieldModulation2023}, or artificial intelligence for diffractive optics \cite{khoninaPerspectiveArtificialIntelligences2024}. Unlike these surveys, this review does not compare beam-shaping technologies as competing components. Instead, it evaluates them within a common hardware–algorithm co-design framework using a~unified seven-axis benchmark.

\newpage 

Section~\ref{sec:methods} sets out the search procedure and the seven-axis frame. Section~\ref{sec:hardware} reviews the hardware classes and compares them on a master table and the refresh-rate--average-power plane. Section~\ref{sec:algorithms} reviews the hologram-generation algorithms. Sections~\ref{sec:micromach} and~\ref{sec:tpp} test the frame against subtractive micromachining and additive two-photon polymerisation, and section~\ref{sec:hybrid} against the hybrid architectures. Section~\ref{sec:challenges} turns the survey into an axis-by-axis verdict and six measurable targets, and section~\ref{sec:conclusions} draws the conclusions and proposes a reporting standard.

\section{Methods}
\label{sec:methods}
The review is built on a documented, repeatable literature search and a single comparison frame
applied to every class, so that both the selection and the comparison can be checked.

\begin{figure}[!h]
  \centering
  \includegraphics[width=0.72\textwidth]{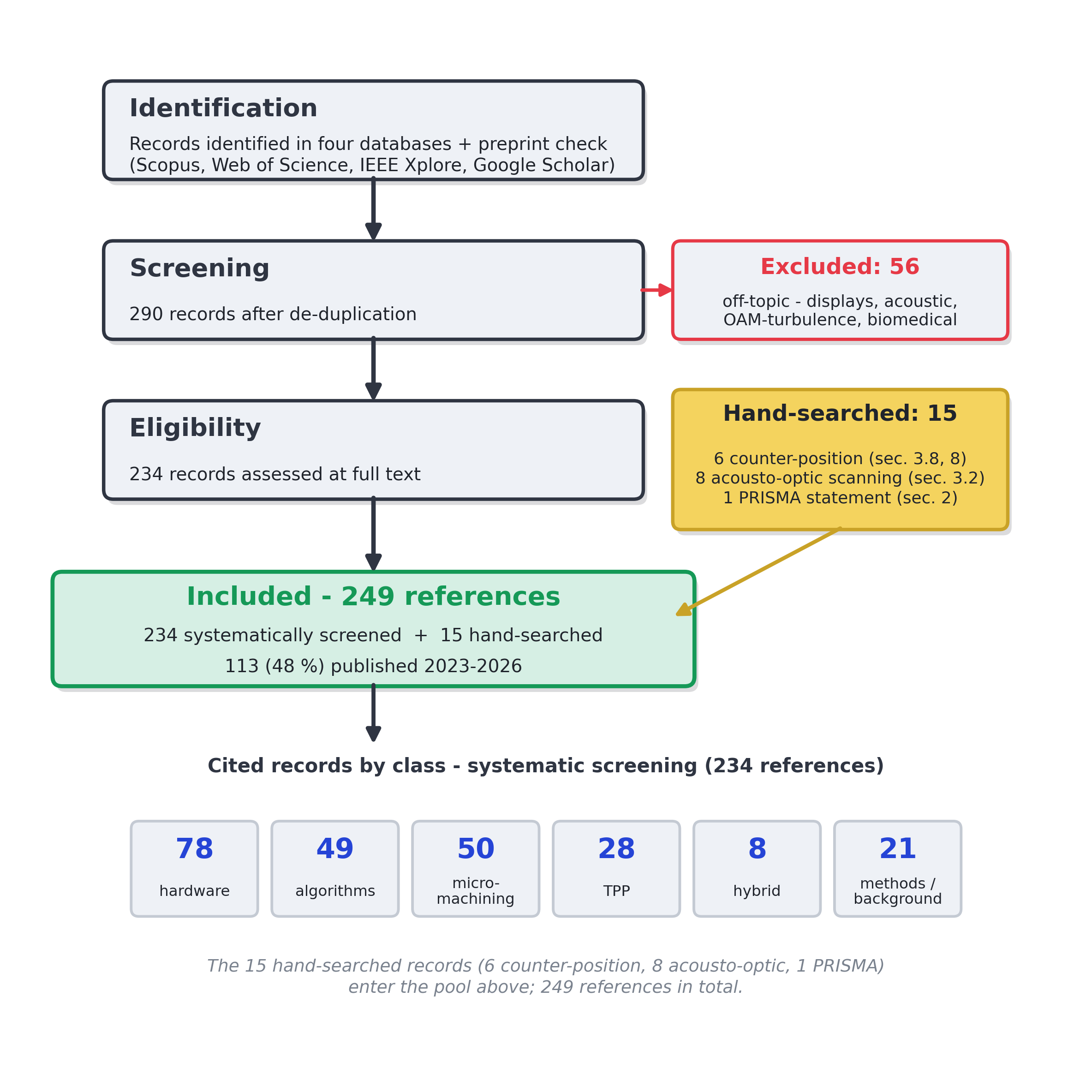}
  \caption{PRISMA-style flow of the literature search. Records from four databases and a preprint check were de-duplicated to 290, then screened to 234 cited sources. A further 15 hand-searched sources bring the total to 249. The lower row groups the 234 cited records by chapter.}
  \label{fig:prisma}
\end{figure}

The literature was screened in Scopus, Web of Science and IEEE Xplore, with Google Scholar as a secondary check for recent preprints. Searches ran from January to July 2026. The query joined three concept groups with Boolean AND: hardware (spatial light modulator, LCoS, DMD, deformable mirror, diffractive optical element, metasurface, freeform optics), algorithm (computer-generated hologram, Gerchberg--Saxton, MRAF, neural network, diffractive neural network) and application (femtosecond, picosecond, two-photon polymerisation, direct laser writing, glass cutting, stealth dicing, surface texturing, waveguide writing, parallel ablation). The primary window was 2018--2026. Older work was kept only where it introduced a method still in use, such as the Gerchberg--Saxton iteration \cite{whyteExperimentalDemonstrationHolographic2005}, the foundational SLM and DOE treatments \cite{efronSpatialLightModulators1989, skerenDesignBinaryPhaseonly2000, shealyTheoryGeometricalMethods2000}, or the early multiphoton-microfabrication reports \cite{maruoRecentProgressMultiphoton2008}. Title-and-abstract and full-text screening were each done by two reviewers, and only English-language records were kept. The screening is PRISMA-informed \cite{pagePRISMA2020Statement2021}, with the flow summarised in figure~\ref{fig:prisma}.

The review cites 249 sources: 234 from the systematic screen, which are the basis for all per-year and per-chapter metrics, plus 15 hand-searched sources (six industrial counter-perspectives, eight acousto-optic scanning references, and the PRISMA guideline). The counter-perspective sources were added deliberately. The search terms favour the programmable hardware literature, so the strongest case for retaining static optics would otherwise be under-represented. That case is engaged directly in sections~\ref{sec:hardware_summary} and~\ref{sec:challenges}. For transparency, two references are the present authors' own work \cite{koblihaSuperhydrophobicSelfcleaningAluminium2025, hauschwitzRapidLaserinducedNanostructuring2025}, cited only for specific experimental figures.

Each class was assessed against seven common axes: (i) reconfigurability, the maximum pattern-update rate; (ii) average-power and peak-fluence handling; (iii) optical efficiency; (iv) beam-quality fidelity, through uniformity, contrast, speckle and $M^{2}$; (v) three-dimensional programmability; (vi) capital cost and integration; and (vii) per-pattern computational cost, the recurring cost of computing each modulator state. The comparison is descriptive rather than statistical. For each class, the best published value on each axis is collected in table~\ref{tab:six_axis_master}. Every value in the application tables (tables~\ref{tab:micromach} and~\ref{tab:tpp}) is taken directly from the cited paper. Where a paper does not report a value, the cell is left as a dash rather than estimated. This uneven reporting across the literature is itself discussed in section~\ref{sec:challenges}. 

Two terms recur throughout. The trade-off has \emph{collapsed} on an axis once a programmable device reaches, and sustains, a throughput or quality that only static optics could reach before. A single-shot record does not count. The trade-off \emph{persists} where the historic coupling still binds. The verdict is assigned axis by axis and regime by regime, and the regime verdicts are consolidated in table~\ref{tab:master_verdict}. The average-power collapse would be falsified if the sustained demonstrations fail to replicate, and it extends to peak fluence only once a programmable device is shown to survive a~stated femtosecond-burst fluence over an equivalent duty cycle (section~\ref{sec:challenges}). The peak-fluence axis is therefore reported as open: neither collapsed by default nor persisting by assumption.

\section{SLM hardware and competing modulators}
\label{sec:hardware}
\label{sec:hardware_intro}

Six hardware classes now compete to program a laser wavefront: liquid-crystal-on-silicon spatial light modulators (LCoS-SLMs), digital micromirror devices (DMDs), deformable mirrors (DMs), diffractive optical elements (DOEs), freeform refractive shapers, and metasurfaces. They differ in refresh rate, power handling, optical efficiency, three-dimensional programmability, and capital cost. Earlier surveys catalogued these trade-offs \cite{khoninaPerspectiveArtificialIntelligences2024, kazanskiyExploringFunctionalCharacteristics2024, harrisonProgressHighpowerHighintensity2024}. The comparison below instead reads all six classes on a seven-axis frame.

\subsection{Operating principles overview}
\label{sec:slm_principles}

\begin{figure}[!htbp]
  \centering
  \includegraphics[width=0.95\textwidth]{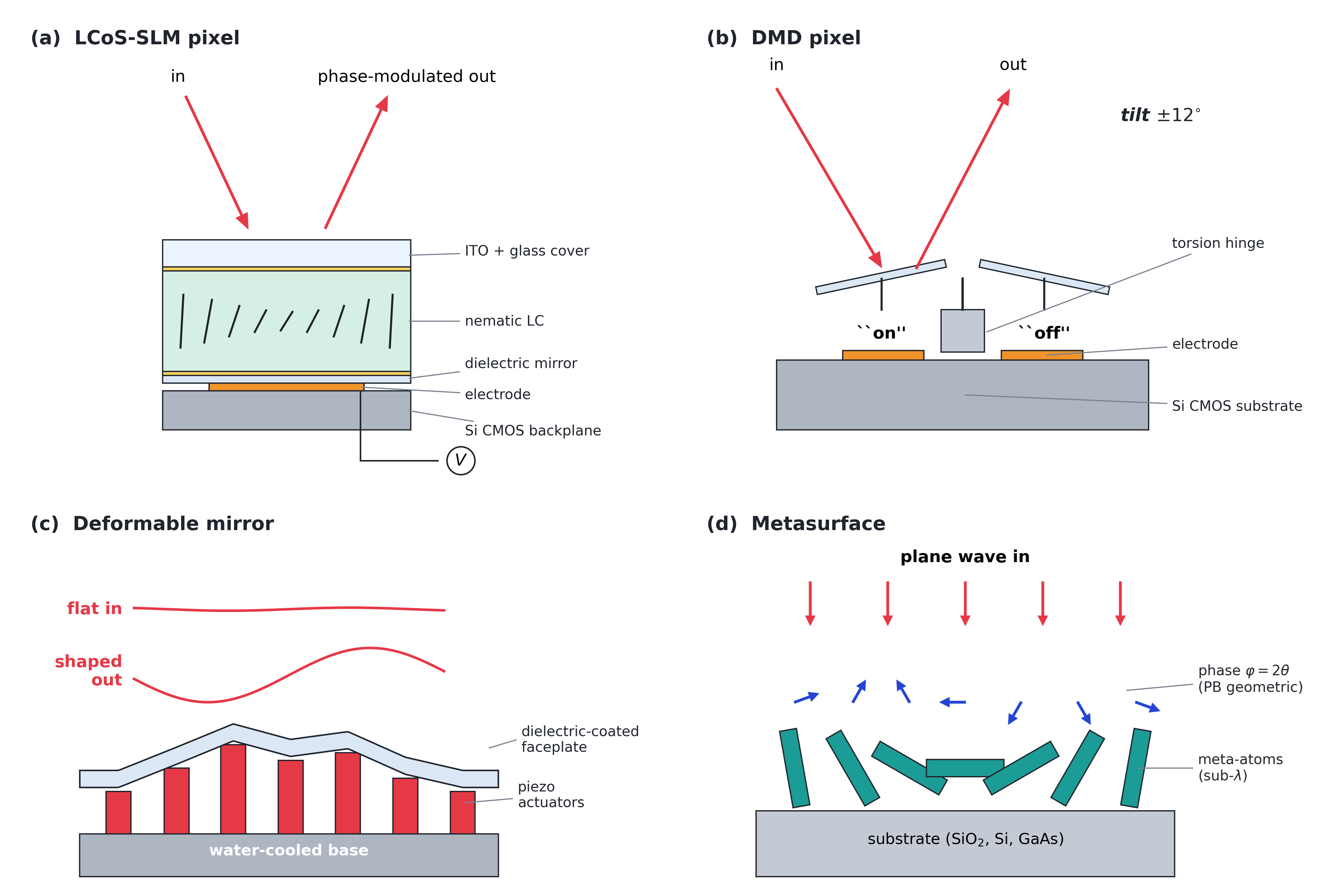}
  \caption{Cross-sections of the four principal hardware classes. (a)~LCoS-SLM: a voltage tilts the nematic-LC director \cite{zhangFundamentalsPhaseonlyLiquid2014, lazarevDisplayPhaseonlyLiquid2019}. (b)~DMD: a bistable torsion-hinge micromirror \cite{hornbeckDeformableMirrorSpatialLight1990, hornbeckDigitalLightProcessing1997}. (c)~Deformable mirror: a~piezo-actuated cooled faceplate \cite{toporovskyWatercooledStackedactuatorFlexible2021, bremerDesignImplementationDynamic2024}. (d)~Metasurface: rotated meta-atoms imparting a~Pancharatnam--Berry phase $\varphi = 2\theta$ \cite{bomzonSpacevariantPancharatnamBerry2002, yuFlatOpticsDesigner2014, khorasaninejadMetalensesVisibleWavelengths2016}.}
  \label{fig:device_cross_sections}
\end{figure}

Each class imprints a pattern through a different mechanism. An LCoS-SLM rotates its liquid-crystal director with an applied voltage. A DMD tilts an array of micromirrors. A deformable mirror flexes a continuous surface. A DOE or freeform shaper uses a fixed surface relief. And a~metasurface imposes a geometric or resonant phase through sub-wavelength meta-atoms. Figure~\ref{fig:device_cross_sections} shows the four principal cross-sections \cite{goodmanIntroductionFourierOptics2017, zhangFundamentalsPhaseonlyLiquid2014, hornbeckDeformableMirrorSpatialLight1990, yuFlatOpticsDesigner2014, khoninaAdvancementsApplicationsDiffractive2024}.

Despite this diversity, one description unifies them. Each is a thin phase- or amplitude-screen that multiplies the input field, $U_{\rm out}(x,y) = t(x,y)\,U_{\rm in}(x,y)$, and the same Fresnel integral then propagates the result to the focal plane. The classes differ only in the range of $t(x,y)$ they synthesise, how fast they reprogram it, and how much average power and peak fluence they survive \cite{goodmanIntroductionFourierOptics2017}. \newpage

Acousto-optic deflectors and modulators (AODs/AOMs) sit outside this taxonomy. They reprogram faster than any shaper here, giving microsecond, inertia-free deflection at megahertz rates \cite{kastelikDoubleAcoustoopticDeflector2017, antonovAcoustoopticDeflectorHigh2018}, but they steer the beam rather than shape its wavefront. Their natural role is as the fast partner to a galvanometer scanner. Scanners that combine an acousto-optic deflector with a galvanometer reach beam speeds of $10^{2}$ to $10^{3}$~m~s$^{-1}$ \cite{franzCharacterizationHybridScanning2022} and cut the time to drill hole arrays or to chamfer edges by up to about 80~\% \cite{springerUltrashortPulseLaser2024, fengHighefficiencyLaserProcessing2026, liangLaserHighefficientChamfering2026}. Acousto-optic holography has recently gone further, imprinting quasi-2D patterns that switch from pulse to pulse and splitting a femtosecond beam into individually addressable bottle beams \cite{akemannAcoustoopticHolographyPseudotwodimensional2024, obydennovIndependentMulticoloredBottlebeam2023}. Even so, it works only as a serial device with a small aperture and low efficiency.

\subsection{Liquid-crystal-on-silicon SLMs}
\label{sec:lcos}

The phase-only LCoS-SLM uses the field-controlled birefringence of a nematic liquid-crystal layer on a silicon CMOS backplane. A per-pixel voltage tilts the local director, changing the extraordinary index seen by light polarised along the rubbing axis. Each pixel therefore acts as a tunable retarder spanning $[0, 2\pi]$ at the design wavelength \cite{zhangFundamentalsPhaseonlyLiquid2014, lazarevDisplayPhaseonlyLiquid2019}. The same voltage-driven index change underlies the photorefractive light valves of Bortolozzo \etal \cite{bortolozzoBeamCouplingPhotorefractive2008} and the multiple-quantum-well Fabry--P\'erot devices of Ahearn \etal \cite{ahearnMultipleQuantumWell2001}.

The LCoS-SLM is the principal instrument for programmable beam shaping and offers the most flexible modulation of any class \cite{efronSpatialLightModulators1989, yangReviewLiquidCrystal2023, xuLightFieldModulation2023}. Modern devices reach $\sim 10^6$ phase pixels at 3--12.5~\textmu m pitch and refresh up to $\sim 720$~Hz \cite{buskeHighFidelityLaser2023, lazarevDisplayPhaseonlyLiquid2019}. That ceiling is physical: the nematic director relaxes on a millisecond timescale set by rotational viscosity and elastic torque. This caps the phase-update rate below the kilohertz \cite{zhangFundamentalsPhaseonlyLiquid2014}. For the nematic device, the open lever is therefore pattern-computation time, not display time, which is why the algorithmic advances of Section~\ref{sec:algorithms} matter.

Average-power tolerance is the LCoS-SLM's defining limit. The silicon backplane absorbs the light the mirror coating does not reflect, and the heat distorts phase and eventually damages the liquid crystal \cite{buskeAdvancedBeamShaping2022, buskeHighFidelityLaser2023}. For most of the past decade, this held commercial devices to a few tens of watts, excluding them from industrial lines running at hundreds of watts \cite{lutzEfficientUltrashortPulsed2021, hofmannDesignMultibeamOptics2020}. As one example, a damage threshold of about ${\sim}2.6$~W~cm$^{-2}$ has been reported for that generation of devices \cite{maxsonAdaptiveElectronBeam2015}. Early large-facility deployments therefore coped by spreading the beam to lower its intensity and by actively cooling the device, rather than trusting the SLM to withstand the load \cite{barczysDeploymentSpatialLight2013, liSpatialBeamShaping2016, liUsingSpatialLight2018}.

\begin{figure}[!htbp]
  \centering
  \includegraphics[width=0.92\textwidth]{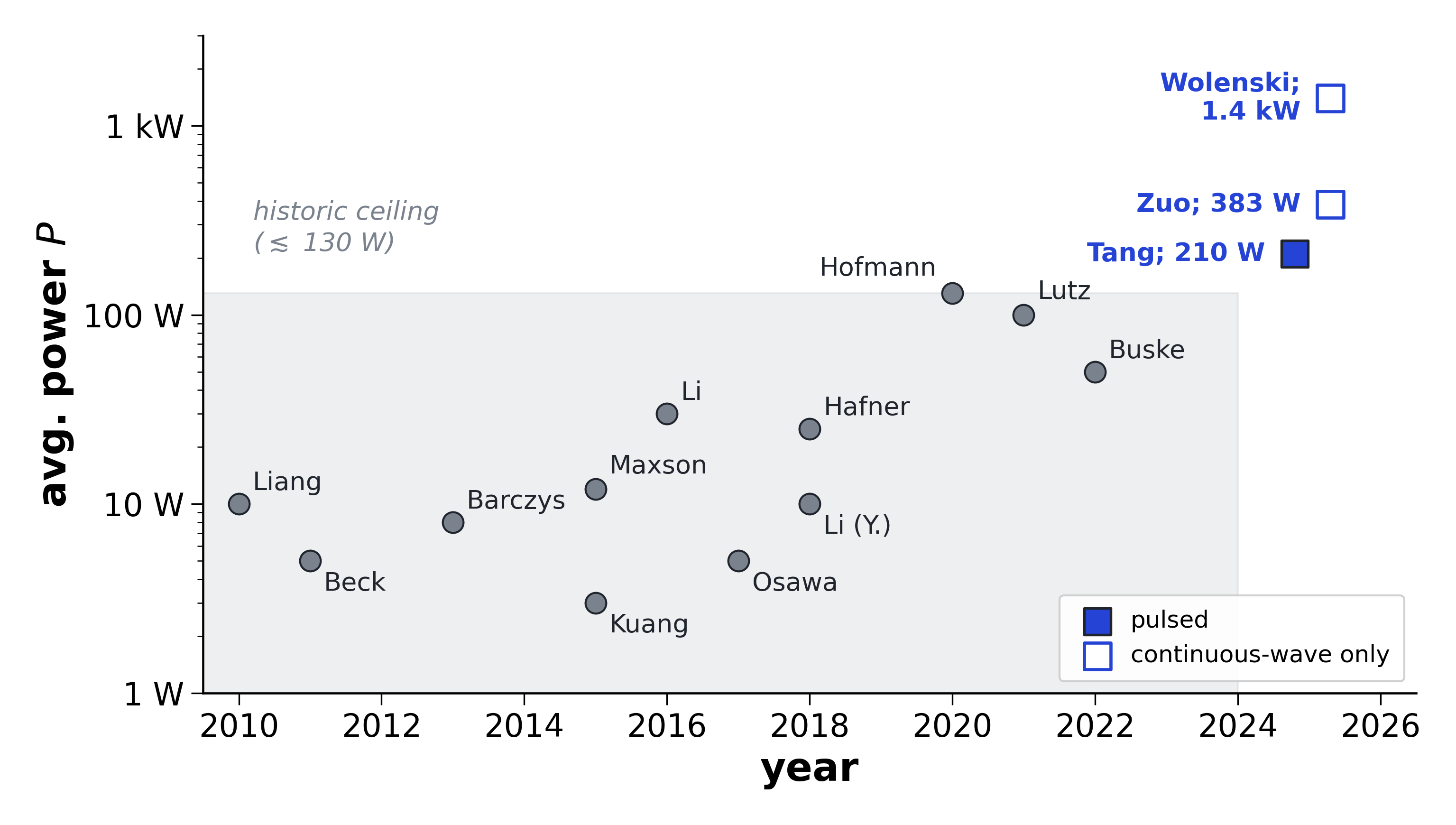}
  \caption{Published average-power ceiling for LCoS-SLM beam shapers, 2010--2025. Grey circles: historic demonstrations under the decade-long $\lesssim 130$~W ceiling \cite{liangHighprecisionLaserBeam2010, beckAdaptiveLaserBeam2011, barczysDeploymentSpatialLight2013, maxsonAdaptiveElectronBeam2015, hofmannDesignMultibeamOptics2020, lutzEfficientUltrashortPulsed2021, buskeAdvancedBeamShaping2022}. Filled square: pulsed record, Tang \etal (210~W) \cite{tangExtendingOperationalLimit2025}; open squares: CW-only Zuo \etal (383~W) \cite{zuoHighPerformanceNIRLaserBeam2025} and Wolenski \etal (1.4~kW) \cite{wolenskiKilowattAveragePower2025}.}
  \label{fig:power_timeline}
\end{figure}

Cooled designs have since raised that ceiling severalfold. Tang \etal \cite{tangExtendingOperationalLimit2025} recovered a full $2\pi$ range at 210~W (2~MHz) using a refractive flat-top generator upstream of a cooled SLM. Zuo \etal \cite{zuoHighPerformanceNIRLaserBeam2025} reported first-order efficiency $\eta = 0.98 \pm 0.01$ at 300~W and a usable phase range beyond $2\pi$ at 383~W under CW illumination, aided by a sapphire window and an optimised dielectric stack. Wolenski \etal \cite{wolenskiKilowattAveragePower2025} described a water-cooled, sapphire-coverglass device operating up to 1.4~kW CW while keeping a full wave of phase, failing only through a reversible transition of the liquid crystal to an isotropic state. Figure~\ref{fig:power_timeline} traces this rise. The pulsed 210~W and CW 383~W points already overturn the objection that SLMs cannot survive industrial average power.

The gains should not be overstated. All three demonstrations rely on customised cooling and are single-group results. An out-of-the-box SLM still runs typically below these results \cite{buskeAdvancedBeamShaping2022, beckAdaptiveLaserBeam2011}. Zuo's own system reports the degree of polarisation falling from 0.97 to 0.69 between 300 and 383~W, an ${\sim}80$~mm thermal focal shift at 300~W, and upstream-optic damage at 407~W \cite{zuoHighPerformanceNIRLaserBeam2025}. The kilowatt result is CW only and silent on peak fluence, and the fs-burst failure mode of LCoS-SLMs remains uncharacterised in the open literature \cite{wolenskiKilowattAveragePower2025, lafargueInVolumeGlassModification2024, balageBesselBeamDielectrics2023}. The field has announced high-power SLMs before: the early-2010s facility deployments read as overcoming the power objection did not transfer beyond their own bespoke cooling. The gain is best read as the frontier being engineered around in specific regimes.

The LCoS-SLM is now routinely built into complete optical systems. For example, combining two SLMs doubles the range of angles the system can address while still refreshing at 30~Hz \cite{kujawinskaLCoSSPATIALLIGHT2012}. The same two-SLM arrangement was later adopted for computer-generated holography in laser processing \cite{sinclairInteractiveApplicationHolographic2004, pangSpecklereducedHolographicBeam2019}. The SLM can also be programmed to correct its own errors. Buske \etal \cite{buskeHighFidelityLaser2023}treated the SLM as a diffractive neural network to compensate astigmatism, crosstalk, and zero-order leakage below 500~\textmu m target distance. Ackermann \etal \cite{ackermannUniformEfficientBeam2021} used both halves of one SLM as a paired amplitude-then-phase shaper for 25~\% uniformity at 73~\% efficiency. And Elsharkawi \etal \cite{elsharkawiBeamShapingNonlinear2026} extended the approach to non-diffracting femtosecond beams with an ${\sim}192$~\textmu m uniform Bessel diameter.

\subsection{Digital micromirror devices}
\label{sec:dmd}

The DMD is the simplest principal class. Each pixel is a 7--14~\textmu m aluminium micromirror on a torsion hinge that toggles between two states in 10--20~\textmu s. The binary tilt makes it an amplitude modulator. Because each mirror is either fully on or fully off, grey levels are built up over time: the mirror is switched on and off rapidly, and the fraction of time it stays on sets the brightness (pulse-width modulation). The cost is that the light arrives as a fast on/off train rather than a steady level, and the timing control is more complex \cite{hornbeckDeformableMirrorSpatialLight1990, hornbeckDigitalLightProcessing1997}.

Its strength is speed and robustness. The pattern-update rate is two orders of magnitude above the LCoS-SLM, with Ren \etal \cite{renTailoringLightDigital2015} reporting 22.7~kHz mask switching on a 44~\textmu s cycle. DMD projection two-photon lithography exploits this directly. Kim \etal \cite{kimRapidPrintingNanoporous2023} printed nanoporous structures at $>0.5$~mm$^2$~s$^{-1}$ per layer with sub-300~nm features \cite{hofmannDesignMultibeamOptics2020}, and G\"ok\c{c}en and Kaya \cite{gokcenExperimentalCharacterizationMultifocal2026} demonstrated real-time reconfigurable multifocal lensing without mechanical refocusing. The DMD is also polarisation-insensitive and works across a broad spectrum \cite{smithNonlinearReconstructionImages2022, kuangUltrafastLaserBeam2015}. This makes it the preferred modulator when the pattern must change on every pulse, as in MHz-burst femtosecond systems, whose only alternative is an external delay-line burst generator \cite{fangPulseBurstGeneration2022}.

Its limit is the binary, amplitude-only nature, which caps diffraction efficiency near 25--50~\% for a generic pattern \cite{renTailoringLightDigital2015, liImagingbasedAmplitudeLaser2016, liUsingSpatialLight2018}. Iterative refinement against camera output recovers part of the loss. Liang \etal  \cite{liangHighprecisionLaserBeam2010} reached 1~\% RMS intensity error, and 0.23~\% after low-pass filtering, over a 1.39~mm$^2$ flat-top at 633 and 1064~nm. Chen \etal  \cite{chenLenslessBeamShaping2026} then suppressed damaging zero-order components with a lensless Fresnel-domain extended-GS formulation. For high-precision phase shaping, the LCoS-SLM remains preferred.

\subsection{Deformable mirrors}
\label{sec:dm}

The DM programs the wavefront by deforming a continuous reflective faceplate rather than addressing pixels. A set of $N$ actuators sets a surface profile $z(x,y) = \sum_{k=1}^{N} a_k\,\phi_k(x,y)$, and because reflection doubles the path difference, the imparted wavefront is $W = 2z$ to first order \cite{cuiLightPeopleProfessor2022, toporovskyWatercooledStackedactuatorFlexible2021}. Its continuous surface avoids the pixel steps and edge diffraction of pixelated devices. The trade-off is a low actuator count (32--127) and, with it, coarse spatial resolution \cite{lefaudeuxNewDeformableMirror2012, samarkinWideApertureBimorphDeformable2022, kasprzackPerformanceThermallyDeformable2013}.

The DM's strength is power handling and efficiency. Bandwidths run from a few hertz in thermal devices to $\geq 2$~kHz in stacked-actuator water-cooled designs \cite{kasprzackPerformanceThermallyDeformable2013, toporovskyWatercooledStackedactuatorFlexible2021, rukosuevRealTimeCorrectionLaser2022}. Moreover, dielectric coatings exceed 99~\% efficiency \cite{chetkinDeformableMirrorCorrection1993, leiDoubledeformablemirrorAdaptiveOptics2012}. Bremer \etal \cite{bremerDesignImplementationDynamic2024} ran a beam-shaping DM at 500~W on a fused-silica faceplate rated for 10~kW, a regime no liquid-crystal device has approached. Deformable mirrors reach even further at petawatt-class facilities. A 320~mm bimorph mirror corrected a 4.2~PW laser and focused it to $1.1 \times 10^{23}$~W~cm$^{-2}$ \cite{lefaudeuxNewDeformableMirror2012, samarkinLargeapertureAdaptiveOptical2022, samarkinWideApertureBimorphDeformable2022, galaktionovLaserBeamPropagation2015}. Such demonstrations mark the high-power regime that the LCoS-SLM is only now approaching.

Its limit is low-order correction. In micromachining, the DM is used mainly for in-process aberration correction and parallelisation \cite{cuiLightPeopleProfessor2022, danielLightFocusingScattering2019, otaEnhancementLaserTrapping2003, nortonLaserGuidestarUplink2014}. The main idea borrowed from astronomy is sensorless adaptive optics. Instead of measuring the wavefront with a Shack--Hartmann sensor, it tunes the correction to maximise an image-quality metric, and so corrects sample-induced aberrations in real time \cite{salterAdaptiveOpticsLaser2019, wangReviewFemtosecondLaser2024, rukosuevRealTimeCorrectionLaser2022}. Section~\ref{sec:hybrid} takes this up again.

\subsection{DOEs, freeform optics and metasurfaces}
\label{sec:doe_meta}

The static class shapes the beam with a fixed element in one of two ways. The diffractive members work within the scalar-diffraction framework \cite{goodmanIntroductionFourierOptics2017}. A DOE is a surface-relief phase mask whose far-field pattern is the Fourier transform of the encoded wavefront (in intensity). The iterative algorithms of Section~\ref{sec:algorithms_iterative} invert this relation, so a fixed DOE is essentially a frozen computer-generated hologram (CGH) \cite{soiferComputerDesignDiffractive2013, wangHybridGerchbergSaxton2017, niuFireworksAlgorithmBased2024}. A metasurface adds two further handles: the Pancharatnam--Berry geometric phase
$\varphi = 2\theta$ \cite{bomzonSpacevariantPancharatnamBerry2002} and the resonant phase of sub-wavelength meta-atoms \cite{yuFlatOpticsDesigner2014}, which together enable multiplexing by wavelength, polarisation and depth \cite{khorasaninejadMetalensesVisibleWavelengths2016, khoninaAdvancementsApplicationsDiffractive2024}. A freeform optic works by refraction instead. A smooth surface redistributes the rays by geometrical mapping, with high efficiency and no diffraction orders \cite{fengSimplifiedFreeformOptics2017, tillkornAnamorphicBeamShaping2018}.

Efficiency separates the static shapers from the programmable ones. DOE efficiency runs 35--75~\% for binary designs \cite{skerenDesignBinaryPhaseonly2000, katzUsingDiffractiveOptical2018, harfoucheComparisonInterferometricDiffractive2014} and above 90~\% for multi-level continuous-phase designs \cite{zhouDesignDiffractivePhase2001, daiImprovedFourierModal2024}. Freeform refractive shapers even exceed 95~\% \cite{shealyTheoryGeometricalMethods2000, fengSimplifiedFreeformOptics2017, osawaBeamShapingSpatial2017}. Both are static but outperform every programmable device on throughput and power, with Tillkorn \etal  \cite{tillkornAnamorphicBeamShaping2018} reaching $>12$~kW and $>40$~W~mm$^{-1}$ in anamorphic line shapers \cite{ hafnerTailoredLaserBeam2018}.

Machine-learning design is now eroding the DOE's static penalty. Buske \etal \cite{buskeAdvancedBeamShaping2022} trained cascaded DOE arrays as a diffractive neural network for misalignment-robust multi-plane shaping at 1.064~\textmu m. Liao \etal \cite{liaoDifferentiableDesignFreeform2024} used a differentiable B-spline representation for freeform DOEs with $\pm 5$~mm defocus tolerance beyond IFTA \cite{khoninaPerspectiveArtificialIntelligences2024}. Dai \etal \cite{daiMicrofabricationTechniqueHighPerformance2025} reached a 0.53~\% experiment-to-simulation discrepancy on silicon.

\begin{figure}[!htbp]
  \centering
  \includegraphics[width=0.72\textwidth]{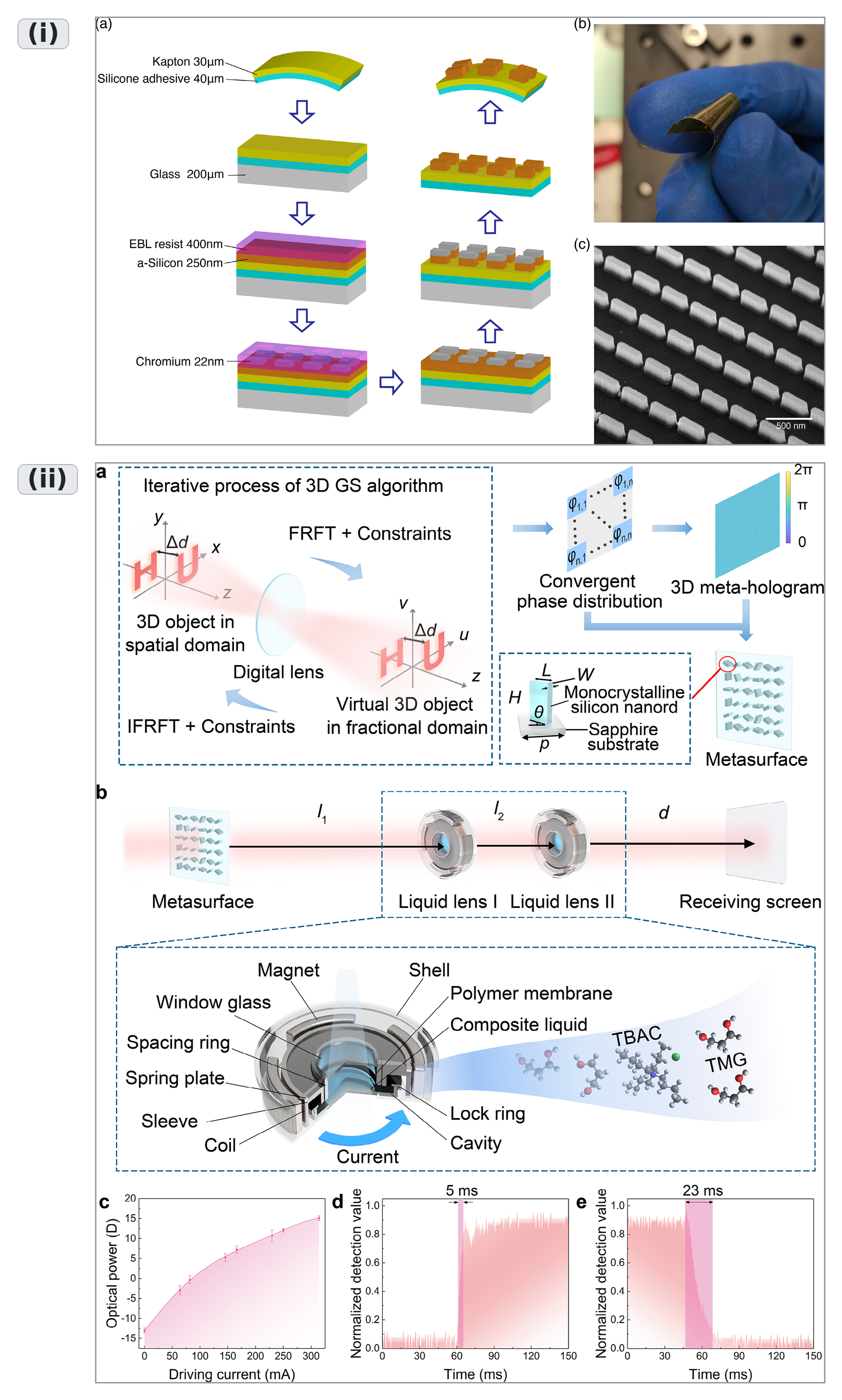}
  \caption{Emerging flat-optics and reconfigurable metasurfaces. (i) Scalable flat optics designed by an evolutionary neural network on a flexible Kapton/glass substrate, shown bent by hand with an SEM of the silicon-nanorod array. (ii) A metasurface cascaded with two liquid lenses giving tunable achromatic 3D meta-holography with a $\sim$5~ms zoom. Panels (i) and (ii) adapted from Makarenko \etal~\cite{makarenkoRobustScalableFlatOptics2021} and Liu \etal~\cite{liuMultiWavelengthAchromatic3D2025}; \textcopyright{} 2021 and 2025 The Author(s), licensed under CC BY~4.0.}
  \label{fig:reuse_devices}
\end{figure}

Like a DOE, a metasurface is a fixed, flat optic, but it packs more into a single layer. One element can encode several functions at once, responding differently to different wavelengths, polarisations, or focal depths \cite{zhaoRecentAdvancesMultidimensional2020, fanHolographicMultiplexingMetasurface2024}. This capability, called multiplexing, is the metasurface's main strength. The best demonstrations so far are at visible wavelengths: a Pancharatnam--Berry device reaching decimetre-scale focal depth \cite{wangDecimeterdepthPolarizationAddressable2024}, and a liquid-lens version giving colour-corrected 3D holography across 473--638~\nm\ with a 5~ms zoom (figure~\ref{fig:reuse_devices}(ii)) \cite{liuMultiWavelengthAchromatic3D2025}. In the deep ultraviolet, flat-top shaping has been shown only in simulation \cite{liUltravioletMetasurfaceenabledFlattop2026}, and infrared evidence is thin. The main infrared example is a single 6.4~mm silicon metalens from Chen \etal \cite{chenLightweightEfficientBeamshaping2025} at 91.8~\% efficiency at 10.6~\textmu m, a CO\textsubscript{2}-laser wavelength far from the micromachining band. Because direct flat-top evidence at ultrafast-micromachining wavelengths is essentially absent, the metasurface is treated here as a static comparator.

Reconfigurable (tunable) metasurfaces are still a research direction, and each gains tunability at a cost, either a very small aperture or an external drive to switch it. Optically pumped AlGaAs devices give a brief phase shift only under a laser pump of 70--180~\textmu J~cm$^{-2}$ \cite{crottiGiantUltrafastDichroism2024}. Thermo-optic and organic devices trade aperture for on-chip addressing, reaching a 0.8$\lambda$ phase stroke over 500--2600~\nm\ \cite{barlandReconfigurableDesignThermooptically2023} or two-channel holography without per-pixel electrodes \cite{chenLightdrivenPhaseTransition2024, huangOrganicMetasurfacesContrasting2025}. Han \etal \cite{hanFemtosecondLaserNonDiffractingBeam2026} instead used a femtosecond quasi-Bessel beam to write a tunable phase-change metasurface (Ge$_2$Sb$_2$Te$_5$) with 9~\nm\ features. The binding limits are survivable power and size. The femtosecond-burst damage threshold stays below that of conventional dielectrics, and the largest optical metasurfaces are around 1~cm$^2$ \cite{lengMetasurfaceMirrorsBased2024, fraserSiliconFresnelZone2023}. Both are improving, for instance with an evolutionary-neural-network design that reaches roughly 80~\% polarisation efficiency on flexible substrates, even under bending (figure~\ref{fig:reuse_devices}(i)) \cite{makarenkoRobustScalableFlatOptics2021}.\newpage

 \newpage

\subsection{Comparison and cost}
\label{sec:hardware_summary}

Table~\ref{tab:six_axis_master} consolidates the best published figure on each of the seven axes for the four programmable classes and the three static ones \cite{khoninaPerspectiveArtificialIntelligences2024, kazanskiyExploringFunctionalCharacteristics2024}. Each cell is a best-in-class value from whichever demonstration reported the highest, so a row is a composite envelope rather than a single purchasable device.

The classes divide cleanly. The LCoS-SLM leads on three-dimensional programmability and arbitrary-pattern fidelity, and with cooling it now reaches 1.4~\kW\ continuous-wave. It remains, however, the lowest-throughput of the programmable classes \cite{buskeHighFidelityLaser2023, pangSpecklereducedHolographicBeam2019, sinclairInteractiveApplicationHolographic2004, wolenskiKilowattAveragePower2025, tangExtendingOperationalLimit2025, zuoHighPerformanceNIRLaserBeam2025, lutzEfficientUltrashortPulsed2021, hofmannDesignMultibeamOptics2020, xuLightFieldModulation2023}. The DMD wins on refresh and polarisation insensitivity at the cost of phase resolution and efficiency \cite{renTailoringLightDigital2015, smithNonlinearReconstructionImages2022, kimRapidPrintingNanoporous2023, kuangUltrafastLaserBeam2015, liUsingSpatialLight2018}. The DM wins on power and efficiency at the cost of pixel count \cite{bremerDesignImplementationDynamic2024, toporovskyWatercooledStackedactuatorFlexible2021, lefaudeuxNewDeformableMirror2012, galaktionovLaserBeamPropagation2015, nortonLaserGuidestarUplink2014, kasprzackPerformanceThermallyDeformable2013}. The DOE and freeform shapers win on throughput, power, and capex but are static \cite{fengSimplifiedFreeformOptics2017, skerenDesignBinaryPhaseonly2000, tillkornAnamorphicBeamShaping2018}. The metasurface is best read as a compact static shaper with a future option of reconfigurability \cite{crottiGiantUltrafastDichroism2024, chenLightdrivenPhaseTransition2024, barlandReconfigurableDesignThermooptically2023, lengMetasurfaceMirrorsBased2024, chenLightweightEfficientBeamshaping2025, fraserSiliconFresnelZone2023}.

The tabulated capex captures only the bare device and is a poor proxy for total cost of ownership. Three further terms dominate industrial selection: the recurring compute to generate each pattern, the amortisation of a one-off fabrication mask, and the cost of retooling when part geometry changes. The decisive variable is therefore production volume and product mix. At high volume and fixed geometry a static DOE or freeform shaper wins, its mask cost amortised to near zero \cite{hafnerTailoredLaserBeam2018, tillkornAnamorphicBeamShaping2018, kazanskiyExploringFunctionalCharacteristics2024}. At low volume or high mix, the programmable device wins by avoiding retooling \cite{lutzEfficientUltrashortPulsed2021}. Let $C_p$ and $C_s$ be the programmable and static acquisition costs, $r$ the retooling cost per geometry change, and $n$ the number of changes over the run. The static shaper is cheaper while:
\begin{equation}
  C_s + n\,r \;<\; C_p,
  \qquad\text{that is, for}\qquad
  n \;<\; n^{\ast} \;\equiv\; \frac{C_p - C_s}{r},
  \label{eq:breakeven}
\end{equation}
\noindent and the programmable device becomes competitive once $n$ exceeds the break-even count $n^{\ast}$. The surveyed literature reports none of $C_p$, $C_s$ and $r$ together for a single production line, so $n^{\ast}$ cannot yet be evaluated. Order-of-magnitude catalogue figures ($C_p \sim \mbox{EUR}\,20$--$40$k, $C_s \sim \mbox{EUR}\,1$--$5$k, $r \sim \mbox{EUR}\,1$--$3$k) place $n^{\ast}$ in the single-to-low-tens range. Equation~\ref{eq:breakeven} omits the recurring per-reprogramming compute cost $C_{\rm comp}$, which only the programmable device pays and which raises $n^{\ast}$ further toward the static optic. These three quantities are what future industrial-facing work must report together. \vspace{-0.75em}

\begin{figure}[!h]
  \centering
  \includegraphics[width=0.9\textwidth]{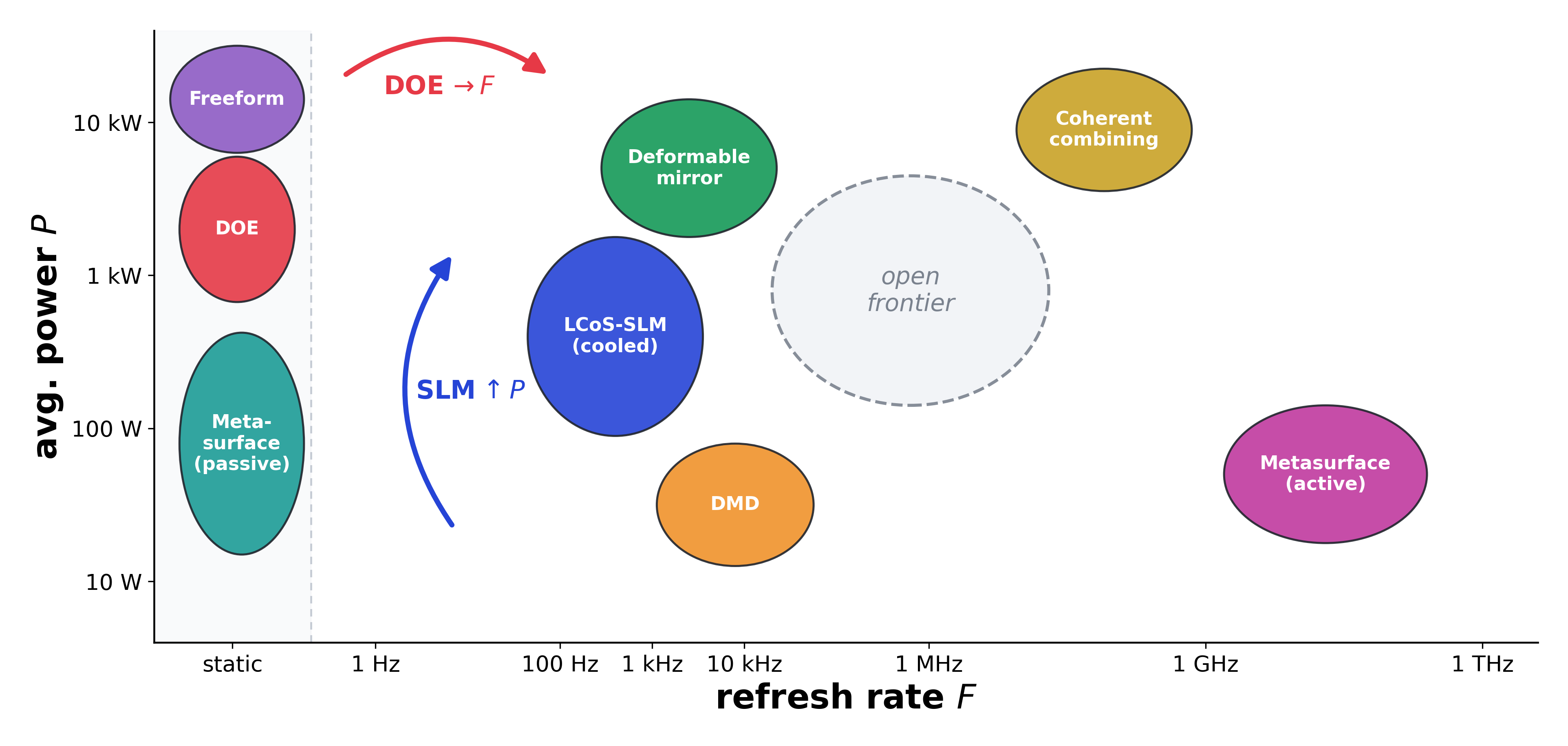}
  \caption{Hardware classes in the refresh-rate--average-power ($F$--$P$) plane. Static ($F=0$) classes fill the left strip, and each bubble marks a published operating envelope. Two recent migrations dominate: LCoS-SLM upward in power \cite{tangExtendingOperationalLimit2025, zuoHighPerformanceNIRLaserBeam2025, wolenskiKilowattAveragePower2025} and DOE upward in refresh via ML re-design \cite{buskeAdvancedBeamShaping2022, liaoDifferentiableDesignFreeform2024, khoninaPerspectiveArtificialIntelligences2024}. The dashed ellipse marks the empty high-$F$, high-$P$ corner, reached only by coherent combining \cite{weberBasicPropertiesHighDynamic2025}.}
  \label{fig:fp_plane}
\end{figure}

\begin{sidewaystable}[H]
  \centering
  \footnotesize
  \setlength{\tabcolsep}{3pt}
  \caption{Master seven-axis comparison of beam-shaping hardware classes. Each entry is the best published figure for that class, not necessarily measured on one device. The axes are defined in section~\ref{sec:methods}, the superscripts refer to the notes below.}
  \label{tab:six_axis_master}
  \begin{tabularx}{\textwidth}{@{}l C C C C L C l l@{}}
    \toprule
    \textbf{Class} &
    \textbf{Refresh $F$} &
    \textbf{Avg.\ power $P$} &
    \textbf{Peak fluence}\textsuperscript{b} &
    \textbf{Opt.\ eff.\ $\eta$} &
    \textbf{Spatial fidelity} &
    \textbf{3D} &
    \textbf{Capex}\textsuperscript{a} &
    \textbf{Compute}\textsuperscript{d} \\
    \midrule
    LCoS-SLM (cooled) &
    $\leq 720$~\Hz~\cite{lazarevDisplayPhaseonlyLiquid2019} &
    $1.4$~\kW\ CW; 210~\W\ pulsed~\cite{wolenskiKilowattAveragePower2025, tangExtendingOperationalLimit2025} &
    n.c. &
    $0.98 \pm 0.01$~\cite{zuoHighPerformanceNIRLaserBeam2025} &
    $\sim 10^6$ phase pixels~\cite{buskeHighFidelityLaser2023} &
    full &
    high &
    per-pattern (GPU) \\
    DMD &
    $\sim 22.7$~\kHz~\cite{renTailoringLightDigital2015} &
    n.c. &
    n.c. &
    $0.25$--$0.50$~\cite{renTailoringLightDigital2015} &
    binary, $\sim 2\times 10^6$ pixels~\cite{renTailoringLightDigital2015} &
    partial &
    moderate &
    per-pattern (GPU) \\
    Deformable mirror &
    $\sim 2$~\kHz~\cite{toporovskyWatercooledStackedactuatorFlexible2021} &
    $0.5$--$10$~\kW~\cite{bremerDesignImplementationDynamic2024} &
    high (dielectric)~\cite{bremerDesignImplementationDynamic2024} &
    $> 0.99$~\cite{toporovskyWatercooledStackedactuatorFlexible2021, bremerDesignImplementationDynamic2024} &
    $32$--$127$ actuators~\cite{lefaudeuxNewDeformableMirror2012, samarkinWideApertureBimorphDeformable2022} &
    partial (stroke-limited) &
    very high &
    low-order \\
    DOE (multi-level) &
    $0$ (static) &
    multi-\kW~\cite{kazanskiyExploringFunctionalCharacteristics2024, khoninaAdvancementsApplicationsDiffractive2024} &
    high (bulk) &
    $> 0.90$~\cite{zhouDesignDiffractivePhase2001, daiImprovedFourierModal2024, daiMicrofabricationTechniqueHighPerformance2025} &
    continuous phase &
    full (fixed at fabrication) &
    low &
    none (fixed) \\
    Freeform refractive &
    $0$ (static) &
    $> 12$~\kW~\cite{tillkornAnamorphicBeamShaping2018} &
    high (bulk) &
    $> 0.95$~\cite{shealyTheoryGeometricalMethods2000, tillkornAnamorphicBeamShaping2018} &
    continuous surface &
    none &
    moderate &
    none (fixed) \\
    Metasurface (passive) &
    $0$ (static) &
    research-scale~\cite{lengMetasurfaceMirrorsBased2024, chenLightweightEfficientBeamshaping2025} &
    68~J\,cm$^{-2}$ (6~\ns)~\cite{wangVortexfieldEnhancementHighthreshold2024} &
    $0.66$--$0.92$~\cite{khorasaninejadMetalensesVisibleWavelengths2016, chenLightweightEfficientBeamshaping2025} &
    sub-$\lambda$ unit cells &
    full (fixed at fabrication) &
    high &
    none (fixed) \\
    Metasurface (active) &
    \GHz--\THz (pump)\textsuperscript{c}~\cite{crottiGiantUltrafastDichroism2024} &
    research-scale &
    n.c. &
    n.c.\ (resonant mechanism) &
    sub-$\lambda$ unit cells &
    partial (research-stage) &
    n/a &
    per-frame (pump) \\
    \bottomrule
  \end{tabularx}

  \smallskip
  \begin{minipage}{\textwidth}
    \footnotesize
    \textsuperscript{a}\,Capex tier ranks acquisition-plus-integration burden, not price: \emph{low}, a commodity or single-mask part; \emph{moderate}, a catalogue component; \emph{high}, an active device needing a driver and cooling; \emph{very high}, a large-aperture actuated system. Research-stage devices are ``n/a''.\\[2pt]
\textsuperscript{b}\,Peak-fluence tolerance under fs-burst loading. ``n.c.''~= not characterised in the open literature for any programmable class (LCoS-SLM, DMD, active metasurface). ``high (bulk)'' and ``high (dielectric)'' denote established static- or dielectric-optic tolerance, not a measured fs-burst value; the passive-metasurface entry is a ns-pulse laser-induced damage threshold (LIDT), unvalidated under fs bursts. The nearest fs-regime measurements still fall short of a usable figure \cite{ramousseFemtosecondDamageThreshold2021, ramousseThermoopticalSLMDamageThreshold2025, schwarzImpactThresholdAssessment2021} and are discussed in section~\ref{sec:challenges}.\\[2pt]
\textsuperscript{c}\,An optically pumped pump-probe relaxation time, not an electronically addressed pattern-update rate, so not directly comparable to the other six classes' $F$ values.\\[2pt]
\textsuperscript{d}\,Recurring compute or energy to generate each modulator state. Static classes incur none once fabricated. Programmable classes pay it on every reprogramming: GPU-backed CGH or learned inference (LCoS-SLM, DMD), a low-order Zernike solve (DM), or a per-frame control-laser exposure (active metasurface). Rarely benchmarked (section~\ref{sec:algo_dl}), so entries are qualitative.
  \end{minipage}
\end{sidewaystable}

The inequality also bounds the present claim. Most micromachining is high-volume and fixed-geometry, such as glass cutting, high-runner marking, and single-geometry texturing. Here the number of geometry changes, $n$, stays well below the break-even count $n^{\ast}$, so a static shaper stays the rational choice no matter how far programmable devices improve. Refractive multi-spot shapers are already used at scale. One supplier reports more than 140 installations in multi-kilowatt welding equipment \cite{laskinBeamShapingMultimode2024, kiddFreeformRefractiveBeamShaping2024, volkertInnovativeBeamShaping2024}. Manufacturing reviews point to cost, complexity, and thermal-damage risk as the main barriers for DOEs and SLMs alike \cite{chenReviewBeamShaping2026, coherentDiffractiveOpticalElementsHighPower2023, holoorSingleElementMultiElement2025}. The collapse documented here therefore widens the range of jobs where programmable optics make sense, but only for high-mix, high-value work where $n$ exceeds $n^{\ast}$: low-volume photonic-device writing, reconfigurable parallel ablation and rapid changeover. Figure~\ref{fig:fp_plane} maps these regions in the refresh-rate--average-power plane.
\section{Hologram-generation algorithms}
\label{sec:algorithms}

\begin{figure}[!htbp]
  \centering
  \includegraphics[width=0.98\textwidth]{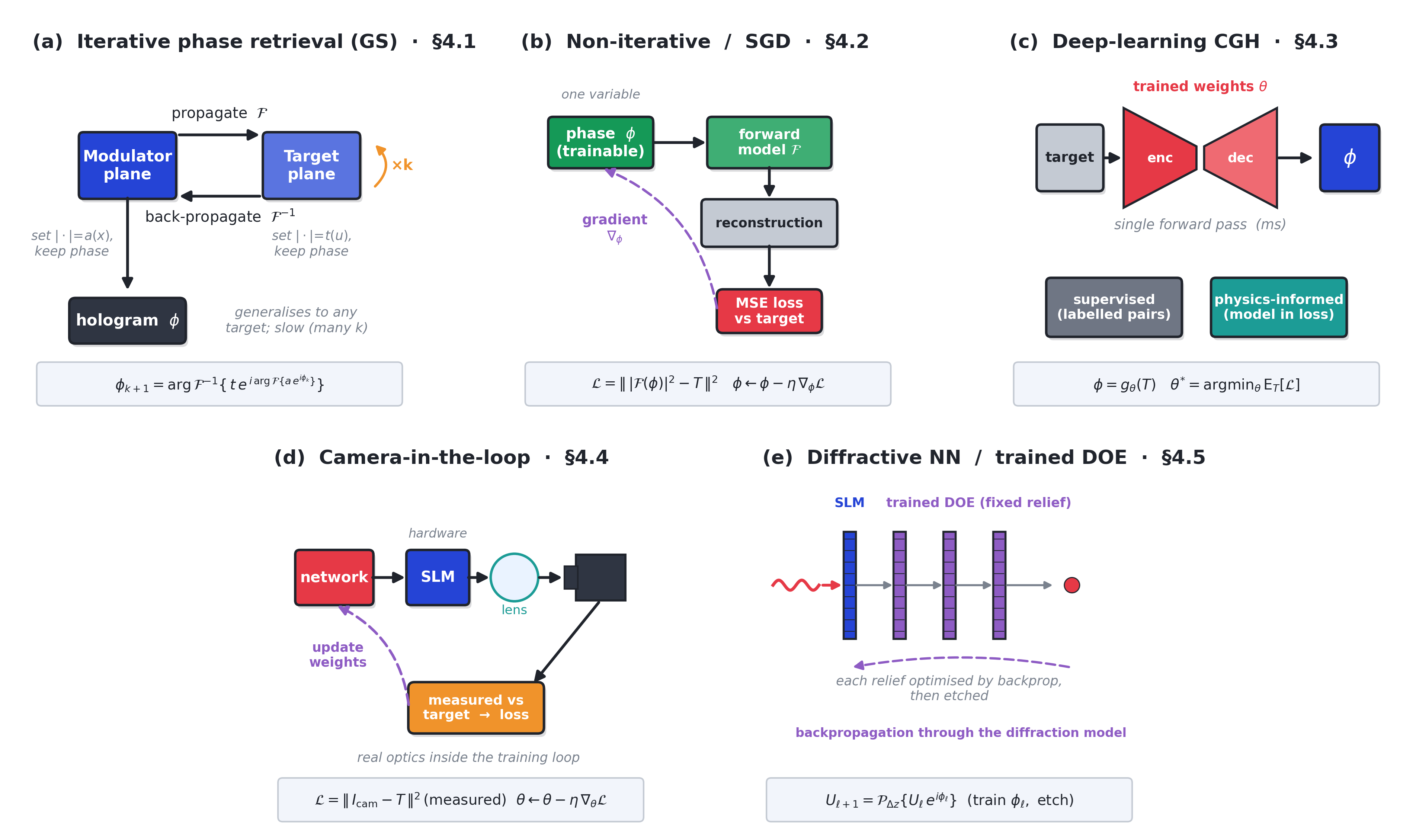}
  \caption{Architectures of the five hologram-generation families. (a) Iterative Gerchberg--Saxton resets the amplitude to the known constraint on each pass between the modulator and target planes. (b) Gradient descent optimises the phase map directly for a single target. (c) Deep-learning CGH infers the phase in one trained pass. (d) Camera-in-the-loop training uses measured reconstructions to drive the loss. (e) A trained DOE optimises its surface relief by backpropagation. All descend from (a).}
  \label{fig:algo_arch}
\end{figure}

The hardware classes of section~\ref{sec:hardware} do not shape a beam on their own. Each needs a control pattern, a phase map for the LCoS-SLM, a mask sequence for the DMD, a Zernike vector for the deformable mirror, or a fixed relief for the DOE, computed by an inverse-design algorithm that maps the target intensity to the modulator state. For most of the past three decades, this algorithmic layer, not the hardware, has set the throughput ceiling of programmable beam shaping \cite{khoninaPerspectiveArtificialIntelligences2024, yuUseDeepLearning2025}. The move to machine-learning inverse design is the main reason the frontier between throughput and flexibility of section~\ref{sec:intro} is now collapsing \cite{yuUseDeepLearning2025, liaoDifferentiableDesignFreeform2024, eybposhDeepCGH3DComputergenerated2020}. This chapter reviews five families of hologram-generation algorithms: iterative phase retrieval, non-iterative and gradient methods, deep-learning holography, camera-in-the-loop systems, and diffractive neural networks.The deformable mirror falls outside this set, because it is driven by a low-order Zernike or influence-function fit rather than a hologram algorithm. It is treated with the hardware in section~\ref{sec:hardware}. Figure~\ref{fig:algo_arch} compares their architectures.

\subsection{Iterative phase retrieval algorithms}
\label{sec:algorithms_iterative}

The Gerchberg--Saxton iteration has been the working algorithm of computer-generated holography for almost five decades. It remains the baseline against which modern methods are benchmarked \cite{gerchbergPracticalAlgorithmDetermination1972, whyteExperimentalDemonstrationHolographic2005, sinclairInteractiveApplicationHolographic2004}. The field is propagated back and forth between the modulator and target planes. At each plane, the computed amplitude is replaced by the known constraint while the phase carries forward, and the cycle repeats until the reconstruction error stalls \cite{whyteExperimentalDemonstrationHolographic2005, pangSpecklereducedHolographicBeam2019}. Whyte and Courtial \cite{whyteExperimentalDemonstrationHolographic2005} and the multi-plane variant of Sinclair \etal \cite{sinclairInteractiveApplicationHolographic2004} established the framework that later work inherits. GS is the right choice whenever a repeatable, auditable pattern is worth the slow computation.

Speckle and reconstruction artefacts are the main practical limit, and the standard remedies fall into three groups. The first, the mixed-region amplitude-freedom (MRAF) family, relaxes the amplitude constraint in a free region around the signal, giving flat-top profiles without edge ringing \cite{pangSpecklereducedHolographicBeam2019, liuDoubleAmplitudeFreedom2022, chenPhaseHologramOptimization2021}. The second, weighted-GS, reweights the per-pixel error as the iteration proceeds. In this area, Wu \etal \cite{wuAdaptiveWeightedGerchbergSaxton2021} gained about 4.8~dB in peak signal-to-noise ratio (PSNR) at a fixed iteration count, though still at multi-second CPU runtimes. The third uses bandwidth limits and digital pre-filters to suppress high-frequency speckle at source \cite{chenSpeckleReductionCombination2014, changSpecklesuppressedPhaseonlyHolographic2015, liSpeckleNoiseSuppression2022}. Beyond these three, a recent fabrication-oriented hybrid by Zuo \etal \cite{zuoHybridOptimizedSpeckle2026} combines GS iteration with gradient-descent loss shaping, cutting speckle contrast by 93.6~\% and raising PSNR from 8.83 to 21~dB under conditions relevant to femtosecond fabrication. In addition, Swan \etal \cite{swanHighfidelityHolographicBeam2024} use optimal transport to generate vortex-free initialisations for phase retrieval, guided by phase-diversity imaging of the incident beam. This complementary route improves the starting point rather than the iteration. \newpage

\begin{figure}[!h]
  \centering
  \includegraphics[width=0.9\textwidth]{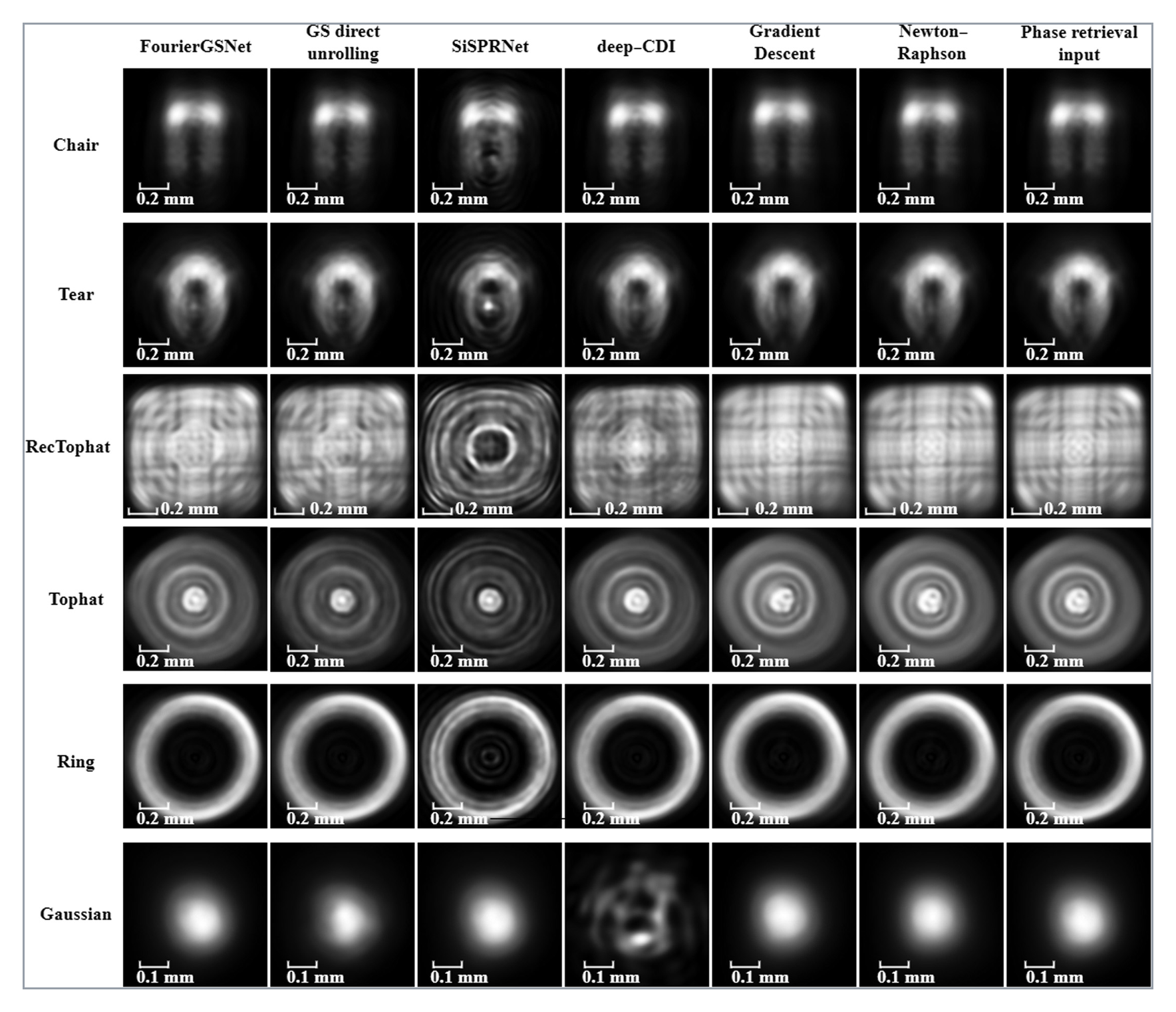}
  \caption{Reconstruction quality of the main phase-retrieval families on shared beam-shaping targets (rows: chair, tear, rectangular and circular flat-top, ring, Gaussian). Columns compare deep-unrolling FourierGSNet against direct GS unrolling, SiSPRNet, deep-CDI, gradient descent, Newton--Raphson, and the raw input, illustrating the accuracy--runtime trade-offs. Adapted from Yan \etal~\cite{yanEfficientGerchbergSaxton2026}; \textcopyright{} 2026 The Author(s), licensed under CC BY~4.0.}
  \label{fig:yan_recon}
\end{figure}

Computational cost is the family's chronic weakness. The earliest three-dimensional solve took several days at $256^3$ voxels on a period workstation \cite{whyteExperimentalDemonstrationHolographic2005}, a runtime incompatible with on-the-fly process control. Two lines of refinement address this: hybrid GS variants that interleave gradient descent with the magnitude update \cite{wangHybridGerchbergSaxton2017, velez-zeaImprovedPhaseHologram2022, zhaiThreedimensionalComputergeneratedHolography2023}, and metaheuristic accelerations such as the fireworks-based GS of Niu \etal \cite{niuFireworksAlgorithmBased2024, nishitsujiFastCalculationComputergenerated2020} and the genetic-algorithm optimisation of Tsai \etal \cite{tsaiApplicationGeneticAlgorithm2015}, all of which stay outside real time. The deep-unrolling FourierGSNet of Yan \etal \cite{yanEfficientGerchbergSaxton2026} narrows the gap, inferring faster than direct GS unrolling while matching or beating its accuracy on high-complexity trains, including full beam-shaping simulations (figure~\ref{fig:yan_recon}). None yet guarantees sustained \kHz\ pattern updates on realistic forward paths, so the gap between GS computation and modulator bandwidth remains an open problem in classical CGH.

Two properties keep GS relevant in an industrial rather than a display setting. It generalises by construction: the same iteration produces a square flat-top by simply changing the target, at no representation cost beyond convergence \cite{whyteExperimentalDemonstrationHolographic2005, sinclairInteractiveApplicationHolographic2004}. Its output is also deterministic and checkable against a stated error metric, so a target yields a reproducible, auditable pattern. That matters for process qualification, where a learned network can fail silently on an out-of-distribution target. The move to non-iterative and learning-based methods has therefore been driven by the demand for real-time updates, not by reconstruction quality or reliability.

\subsection{Non-iterative analytic and gradient-based methods}
\label{sec:algo_noniterative}

A non-iterative algorithm computes the modulator phase in a single deterministic step, removing the per-iteration runtime that dominates GS \cite{velez-zeaNonIterativePhaseOnlyHologram2025, suiNonconvexOptimizationInverse2024}. The most direct route is gradient descent on the phase map: Velez-Zea \cite{velez-zeaNonIterativePhaseOnlyHologram2025} shows a stochastic-gradient pipeline that beats GS on low-contrast targets and matches it on high-contrast ones, so the gain is regime-dependent. A related route treats the phase as a closed-form analytic mapping from the diffraction integral, optionally followed by one global gradient pass \cite{velez-zeaImprovedPhaseHologram2022, wangHybridGerchbergSaxton2017, chenPhaseHologramOptimization2021}. Both suit low-complexity targets such as Gaussian shaping and single-spot generation, where the cost of iteration is unjustified. These methods also differ from the learned methods below: here the computation optimises one hologram for one target and carries no trained model, whereas a network learns its weights once over a target distribution and then infers in a single pass.

\subsection{Deep-learning computer-generated holography}
\label{sec:algo_dl}

Deep learning has been the most active algorithmic development of the past six years. A trained convolutional or generative network outputs the modulator phase for a target in one forward pass, with inference times in milliseconds rather than seconds \cite{eybposhDeepCGH3DComputergenerated2020, yuUseDeepLearning2025}. The cost is up-front training, a representative training distribution, and the loss of GS generality. The foundational 2020--2021 work fixed the reference point. DeepCGH reported a 41~\% accuracy gain over classical CGH at orders-of-magnitude shorter inference on 3D holograms up to 11 megavoxels \cite{eybposhDeepCGH3DComputergenerated2020}. Later variants kept this template and changed only individual pieces: the training loss became adversarial \cite{khanGANHoloGenerativeAdversarial2021}, the cost moved to an offline stage so that inference stays real-time \cite{horisakiThreedimensionalDeeplyGenerated2021}, and the output became binary-amplitude \cite{goiDeeplearningbasedBinaryHologram2020}.

The 2020--2023 wave then mapped one shared deployment problem: each architecture buys progress on one constraint by spending capacity elsewhere \cite{leeDeepLearningBasedFramework2022, chenRealtimeHologramGeneration2024}. Edge-device inference came at the cost of network capacity in Shi \etal's \cite{shiRealtimePhotorealistic3D2021} sub-620-kB CNN at 60~\Hz. Volumetric accuracy came at the cost of compactness in two forms: multi-depth 4K reconstruction from Lee \etal \cite{leeDeepNeuralNetwork2020} and occlusion handling from Yan \etal \cite{yanGeneratingMultiDepth3D2024}. Independence from training-data coverage was achieved by Wu \etal's \cite{wuHighspeedComputergeneratedHolography2021} holoencoder, which embeds diffraction propagation in its decoder. That last move, putting the physics inside the network, defines the physics-informed branch. In these methods, the known forward model is embedded in the loss. The network then supervises itself: it compares its own simulated reconstruction with the target, needing no ground-truth hologram \cite{shuiDiffractionModelinformedNeural2022, zhengDiffractionModeldrivenNeural2023, mashikoExtrapolatedSpecklecorrelationImaging2023, zhengUnsupervisedDeepNeural2024}. Their shared lesson is that the residual error is the imperfect propagation model, not network capacity.

By 2024--2026 the effort had consolidated on three fronts, each closing a specific earlier gap. The first is bridging to real hardware. Yeom \etal \cite{yeomHighqualityPhaseonlyFourier2025} reached 24.89~FPS at $3840 \times 2160$ with a camera-in-the-loop phase-only Fourier network. Hern\'andez-Felipe \etal \cite{hernandez-felipeMultiplaneOptimizingPhase2024} then showed that mundane training choices decide much of the residual multiplane quality. The second is generalisation without retraining. Liu \etal \cite{liuPropagationadaptive4KComputergenerated2025} made the propagation distance a network input, holding 39.25~dB PSNR across a 30~mm depth range. And Zeng \etal \cite{zengDeepLearningFramework2026} suppressed inter-plane crosstalk with a trainable Fourier layer. The third front is the most relevant here, because it targets the fabrication outcome itself rather than a generic hologram. Liu \etal's \cite{liuModelDrivenDeepLearning2026} SMART HoloTPL trains against a broadband polymerisation model and reports speckle-free 3D nanofabrication at 120{,}000~voxels~s$^{-1}$ with 120~\nm\ features. Consolidating reviews and generative variants map the same fronts across the wider catalogue \cite{yuUseDeepLearning2025, shimobabaDeepLearningComputationalHolography2022, kangDeeplearningbasedHologramGeneration2021, yuAsymmetricalNeuralNetwork2023}.

The limitations are now well documented. The training data must represent the target distribution, which is hard for industrial micromachining because the patterns differ widely from one process to the next. Transfer across pixel pitch, wavelength, and depth is incomplete. Unlike the deterministic GS baseline, a network gives no guarantee on a target it was not trained on, and no check that the pattern it produces in one pass has converged \cite{khoninaPerspectiveArtificialIntelligences2024}. Only the latency half of the trade-off can be bounded. A demanding GS or weighted-GS solve runs from hundreds of milliseconds to seconds on CPU \cite{wuAdaptiveWeightedGerchbergSaxton2021}, whereas a trained network infers in tens of milliseconds (for example $\approx 40$~ms at 4K for the camera-in-the-loop network \cite{yeomHighqualityPhaseonlyFourier2025} and $\approx 26$~ms per 4K frame for the propagation-adaptive operator \cite{liuPropagationadaptive4KComputergenerated2025}). The inference gain is thus of order $10^{1}$--$10^{3}$ per pattern, which is why the refresh gain of section~\ref{sec:intro} is real. Energy cost, however, goes unreported: none of these works state joules per pattern, and the GPU-days of training \cite{horisakiThreedimensionalDeeplyGenerated2021} are never divided through the patterns a deployed model will generate \cite{yuUseDeepLearning2025}. These gaps motivate the camera-in-the-loop and physics-informed-DOE directions below.

\subsection{Camera-in-the-loop and end-to-end systems}
\label{sec:algo_citl}

Deep learning assumes the propagation model used for training matches the real optical system, and in practice it does not. Pixel-pitch tolerances, residual aberrations, partial coherence, sample-induced distortion, and SLM non-linearities each introduce a mismatch that synthetic data cannot correct \cite{pengNeuralHolographyCameraintheloop2020, pengSpecklefreeHolographyPartially2021, yooLearningbasedCompensationSpatially2022}. Peng \etal \cite{pengNeuralHolographyCameraintheloop2020} close this by training in a camera-in-the-loop configuration. A camera captures the reconstructed intensity and feeds it into the loss. The reported PSNR rises from 16.1 (GS) and 17.6 (HoloNet open loop) to 19.0 at 1080p. Peng \etal \cite{pengSpecklefreeHolographyPartially2021} latterly extended the method to partially coherent sources, reaching speckle-free reconstruction at three wavelengths. This matters for ultrafast micromachining, where broadened spectra and limited coherence are routine. Yeom \etal \cite{yeomHighqualityPhaseonlyFourier2025} applied the same principle at 24.89~FPS and 4K. The family supplies the empirical correction layer that closes the simulation--experiment gap, at the cost of a calibration step.

\subsection{Diffractive neural networks and physics-informed DOE design}
\label{sec:algo_dnn}

A distinct approach optimises the DOE's physical surface directly. Its surface heights become the trainable parameters, tuned by backpropagation against the same target loss used for a software CGH. There is no intermediate phase map to fabricate from, and once the relief is etched, it stays fixed. Buske \etal \cite{buskeAdvancedBeamShaping2022} demonstrated a diffractive neural network for beam shaping at 1064~\nm\ using $10 \times 10$~mm$^2$ DOEs at 5.56~\um\ pitch. A programmable SLM and a trained DOE can be cascaded, keeping modulator-side reconfigurability while raising DOE-side bandwidth \cite{buskeAdvancedBeamShaping2022, khoninaPerspectiveArtificialIntelligences2024}. Liao \etal \cite{liaoDifferentiableDesignFreeform2024} added a differentiable freeform-DOE framework with a B-spline smoothness constraint that closes the loop between simulation and lithographic patterning. Rahman and Özcan \cite{rahmanIntegrationProgrammableDiffraction2024} generalised this to a programmable-diffraction stack. And Jacob \etal \cite{jacobDynamicBeamShaping2025, liRedefinableNeuralNetwork2024} showed a wavelength-adaptive network producing Gaussian, ring and top-hat profiles on demand at 915, 1064 and 1550~\nm. A wider literature, including surveys of AI across photonics \cite{mahmoudAIdrivenPhotonicsUnleashing2024}, reviews of industrial bottlenecks from manufacturing \cite{murzinArtificialIntelligenceDrivenInnovations2024, murzinComputerScienceIntegrations2024, murzinDigitalEngineeringPhotonics2024}, a foundational reference on deep learning for computational imaging \cite{barbastathisUseDeepLearning2020}, and a perspective linking AI to the design of diffractive optics \cite{khoninaPerspectiveArtificialIntelligences2024}, places this work in context.

The five families are one connected genealogy, rooted in the GS iteration and migrating through its non-iterative, learning-based, camera-in-the-loop and trained-DOE descendants (figure~\ref{fig:algo_arch}). The trained-DOE branch re-enters the hardware discussion of section~\ref{sec:doe_meta}. This algorithmic migration and the hardware migrations of section~\ref{sec:hardware} are two sides of the same shift. Sections~\ref{sec:micromach} and~\ref{sec:tpp} each pick an operating point on this joint design space and test it against a concrete process: the high-power multi-spot corner in subtractive micromachining, and the high-density volumetric corner in additive two-photon polymerisation. In both, the decisive question is which modulator--algorithm pairing survives the process load while reprogramming fast enough to matter.

\section{Beam shaping in ultrafast laser micromachining}
\label{sec:micromach}

Ultrafast laser micromachining is where the hardware and algorithm toolkits of sections~\ref{sec:hardware} and~\ref{sec:algorithms} matter most. The nonlinear absorption of femtosecond and picosecond processing ties feature size, depth, and throughput to the intensity distribution at the focus, so the beam profile is the main lever on the process \cite{jiaRecentProgressFemtosecond2023}. Two scaling laws set the stakes. First, the ablated depth per pulse follows $d(F) \approx \delta\,\ln(F/F_{\rm th})$, where $F$ is the local fluence and $F_{\rm th}$ the ablation threshold. The specific removal rate therefore peaks at an optimum fluence a few times threshold, which for a Gaussian beam is $F_{\rm opt} = e^{2} F_{\rm th}$ \cite{raciukaitisUltraShortPulseLasers2021, shinReviewHighprecisionFemtosecond2024}. Because a Gaussian focus holds only a thin annulus near $F_{\rm opt}$, most of the illuminated area ablates inefficiently, and a flat-top or task-matched profile that keeps the whole area near $F_{\rm opt}$ becomes the single largest lever on efficiency \cite{hafnerTailoredLaserBeam2018, lutzEfficientUltrashortPulsed2021}. Second, heat accumulation sets a limit of its own. When the pulse interval falls below the heat-diffusion time, energy builds up between pulses and the heat-affected zone widens, unless the deposition is shaped against it in space and, increasingly, in time \cite{shinReviewHighprecisionFemtosecond2024, schmidtDynamicBeamShaping2024}. Against this backdrop, four sub-classes now depend on programmable shaping: glass cutting and stealth dicing, surface texturing, photonic-device writing, and multi-spot parallel ablation. Table~\ref{tab:micromach} collects the demonstrations discussed below.

\begin{table}[!htbp]
  \centering
  \footnotesize
  \setlength{\tabcolsep}{4pt}
  \caption{Ultrafast-micromachining demonstrations discussed in this section, grouped by process. Values are quoted from the cited work. A dash (--) marks an unstated quantity. The ``Key result'' column reports each study's own figure of merit in its original unit and is not comparable across rows. The \mbox{Feat./$\lambda$} column gives the smallest reported in-plane feature divided by the wavelength. It allows a rough comparison between studies, not an exact one.}
  \label{tab:micromach}
  \begin{tabularx}{\textwidth}{@{}
    >{\hsize=1.00\hsize\raggedright\arraybackslash}X
    >{\hsize=0.95\hsize\raggedright\arraybackslash}X
    >{\hsize=1.00\hsize\raggedright\arraybackslash}X
    >{\hsize=0.45\hsize\centering\arraybackslash}X
    >{\hsize=0.75\hsize\raggedright\arraybackslash}X
    >{\hsize=0.85\hsize\centering\arraybackslash}X
    c
    >{\hsize=2.00\hsize\raggedright\arraybackslash}X
  @{}}
    \toprule
    \textbf{Study} &
    \textbf{Process} &
    \textbf{Material} &
    \textbf{$\lambda$} &
    \textbf{Pulse} &
    \textbf{Avg.\ power / rep-rate} &
    \textbf{Feat./$\lambda$} &
    \textbf{Key result} \\
    \midrule
    Cheng \etal~\cite{chengFlexibleTunedMultifocus2024} &
    Glass cutting / stealth dicing &
    JGS3 quartz &
    1064~\nm &
    12~\ps, 440~\textmu J &
    20~\kHz &
    -- &
    10 coaxial foci, 60~\um\ spacing, $\sim 1$~\um\ roughness \\
    Qiao \etal~\cite{qiaoFineOptimizationAberration2024} &
    Glass cutting / stealth dicing &
    Si wafer (400~\um) &
    1085~\nm &
    15~\textmu J &
    2.5~\W, 500~mm~s$^{-1}$ &
    -- &
    58.7~\um\ crack, 38.4\% improvement \\
    Wang \etal~\cite{wangIntegratedLCOSSLMBasedLaser2025} &
    Glass cutting / stealth dicing &
    N-type SiC &
    1064~\nm &
    1~\ns &
    0.3~\W, 50~\kHz &
    -- &
    Focal depth 45~\um\ $\rightarrow$ 15~\um \\
    Balage \etal~\cite{balageBesselBeamDielectrics2023} &
    Glass cutting / stealth dicing &
    Glass &
    1030~\nm &
    500~\fs, 1.28~\GHz\ burst &
    -- &
    -- &
    Through-thickness cuts up to 1~mm \\
    Lafargue \etal~\cite{lafargueInVolumeGlassModification2024} &
    Glass cutting / stealth dicing &
    Sodalime, fused silica &
    1030~\nm &
    500~\fs &
    up to 100~\W &
    6.1 &
    6.3~\um\ spot, GHz-burst process window \\
    Ackermann \etal~\cite{ackermannSpotArraysUniform2023} &
    Surface texturing &
    Si (monocrystalline) &
    1064~\nm &
    12~\ps, $\sim 75$~\textmu J &
    20~\kHz, $\sim 600$~\Hz\ update &
    -- &
    $1920 \times 1080$ SLM, uniform spot array \\
    Hauschwitz \etal~\cite{hauschwitzRapidLaserinducedNanostructuring2025} &
    Surface texturing &
    AISI 316L steel &
    1030~\nm &
    2~mJ, 31~J~cm$^{-2}$ &
    50~\kHz &
    0.61 &
    $500 \times 30$~\um\ line, 630~\nm\ LIPSS, 99.88\% reduction \\
    Qiu \etal~\cite{qiuAdaptiveBeamshapingEnabled2025} &
    Surface texturing &
    SiC (single-crystal) &
    520~\nm &
    3--6~\textmu J &
    100~\textmu m~s$^{-1}$ &
    19 &
    $< 0.5$~\um\ RMS error, 10~\um\ grooves \\
    Kawaguchi \etal~\cite{kawaguchiFemtosecondVectorVortex2023} &
    Surface texturing &
    Tungsten &
    520~\nm &
    430~\fs &
    500~\kHz &
    -- &
    Vortex beam broadens LIPSS window \\
    He \etal~\cite{heSlitBeamShaping2022} &
    Waveguide / photonic writing &
    Fused-silica fibre core &
    513~\nm &
    39--125~nJ &
    1~\kHz\ (write rate) &
    2.1 &
    FBG, 1.07~\um\ pitch, $> 30$~dB contrast \\
    Liu \etal~\cite{liuFabricationSinglemodeCircular2021} &
    Waveguide / photonic writing &
    Fused silica (JGS1) &
    1030~\nm &
    270~\fs &
    1~\kHz--1~\MHz &
    9.7 &
    10~\um\ optofluidic channels, $\sim 1$~mm~s$^{-1}$ \\
    Hofmann \etal~\cite{hofmannDesignMultibeamOptics2020} &
    Parallel ablation &
    -- &
    -- &
    -- &
    -- &
    -- &
    $8 \times 8 = 64$ spots, 24--55~\um\ error (multi-beam optics design study; material, $\lambda$ and power not reported) \\
    Lutz \etal~\cite{lutzEfficientUltrashortPulsed2021} &
    Parallel ablation &
    Stainless steel (X5CrNi18-10) &
    1030~\nm &
    0.2--0.4~J~cm$^{-2}$ &
    100~\W &
    -- &
    20 spots, 6~mm$^3$~min$^{-1}$ (0.06~mm$^3$~min$^{-1}$~\W$^{-1}$) \\
    Li \etal~\cite{liHighQualityMicropatternPrinting2024} &
    Parallel ablation &
    SZ2080 photoresist &
    800~\nm &
    35~\fs &
    1~\kHz, 20~mW &
    0.79 &
    10--50~\um\ structures, 0.63~\um\ resolution \\
    Jacob~\cite{jacobPlanarLightValve2024} &
    Parallel ablation &
    Stainless steel (SS304) &
    1064~\nm &
    10~\ps, $\sim 1$~mJ &
    100~\kHz &
    9.4 &
    1088-pixel PLV, $> 30\times$ throughput \\
    \bottomrule
  \end{tabularx}
\end{table}

\subsection{Glass cutting and stealth dicing}
\label{sec:micromach_glass}

Stealth dicing modifies a transparent dielectric in-volume, then cleaves it along the modified plane. It is the established industrial route for cutting silicon, glass, and silicon-carbide wafers and avoids the chipping, debris, and kerf of saw-blade dicing \cite{dudutisIndepthComparisonConventional2020, kumkarComparisonDifferentProcesses2014, kimStudyGlassTGV2023, pecholtReviewLaserMicroscale2011}. The single-focus version is limited by the depth a given numerical aperture can reach, by spherical-aberration loss at depth, and by its serial focal trajectory. Multi-focus SLM implementations relax all three at once. Wang \etal \cite{wangIntegratedLCOSSLMBasedLaser2025} applied integrated LCoS-SLM aberration correction to N-type SiC, cutting the in-volume focal depth from 45 to 15~\um\ and removing the multi-layer damage that has historically constrained SiC slicing (figure~\ref{fig:reuse_micromach}(i)). Qiao \etal \cite{qiaoFineOptimizationAberration2024} corrected aberration directly, improving dicing performance by 38.4~\% and reaching a 58.7~\um\ crack at 2.5~\W\ and 500~mm~s$^{-1}$.

Beyond the wavefront correction above, pulse-format engineering is a second lever. A Bessel beam, formed using a refractive axicon or an SLM-imposed conical phase, decouples the focal length from the numerical aperture and lets the full thickness be modified in a single shot \cite{bhuyanHighAspectRatio2010, guoFemtosecondLaserBessel2022, qinUltrafastLaserProcessing2024, moriSideLobeSuppression2015, osbildSubmicrometerSurfaceStructuring2021}. Lafargue \etal \cite{lafargueInVolumeGlassModification2024} compared the single-pulse, MHz-burst and GHz-burst regimes at 1030~\nm\ and up to 100~\W. Only the GHz burst gave a process window broad enough for high-throughput work in the bulk of sodalime and fused silica. The areal roughness fell from $S_a = 0.42$ to $0.27$~\um\ (figure~\ref{fig:reuse_micromach}(ii)). Balage \etal \cite{balageBesselBeamDielectrics2023} used the same Bessel beam in GHz bursts to cut glass up to 1~mm thick, and adjusting the number of foci further smooths the diced surface \cite{chengFlexibleTunedMultifocus2024}. The current open problems are in-process crack monitoring for closed-loop control \cite{wangReviewFemtosecondLaser2024, wangInsituRealtimeMonitoring2024}, aberration-correction libraries for wafers of arbitrary thickness and index \cite{liaoDifferentiableDesignFreeform2024}, and integrating multi-focus dicing with 100~\W-class sources \cite{zuoHighPerformanceNIRLaserBeam2025, tangExtendingOperationalLimit2025}.

\subsection{Surface texturing}
\label{sec:micromach_texturing}

Surface texturing shows the largest throughput gains from programmable shaping \cite{ackermannSpotArraysUniform2023, hauschwitzRapidLaserinducedNanostructuring2025, schilleHighRateLaserSurface2020}. The area to be textured grows with the square of its size, and a single serial spot cannot keep up. Splitting the beam into an SLM array of $10^2$ to $10^3$ spots multiplies throughput by the same factor, limited only by the energy the source can put into each spot. Ackermann \etal \cite{ackermannSpotArraysUniform2023} ran a $1920 \times 1080$ SLM at a 20~\kHz\ pulse rate, updating the spot pattern at nearly 600~\Hz\ with a clear route to 10~\kHz. Array uniformity suffers a little, but the pattern can be reprogrammed, which a fixed DOE cannot do. The same speed-up carries over to longer pulses. Ahuir-Torres \etal \cite{ahuir-torresNanosecondPulsedLaserBeam2026} used a liquid-crystal SLM to write arbitrary microtextures over a $10 \times 10$~mm steel area in about 2~s, roughly 500 times faster than a serial scan.

\begin{figure}[!h]
  \centering
  \includegraphics[width=0.85\textwidth]{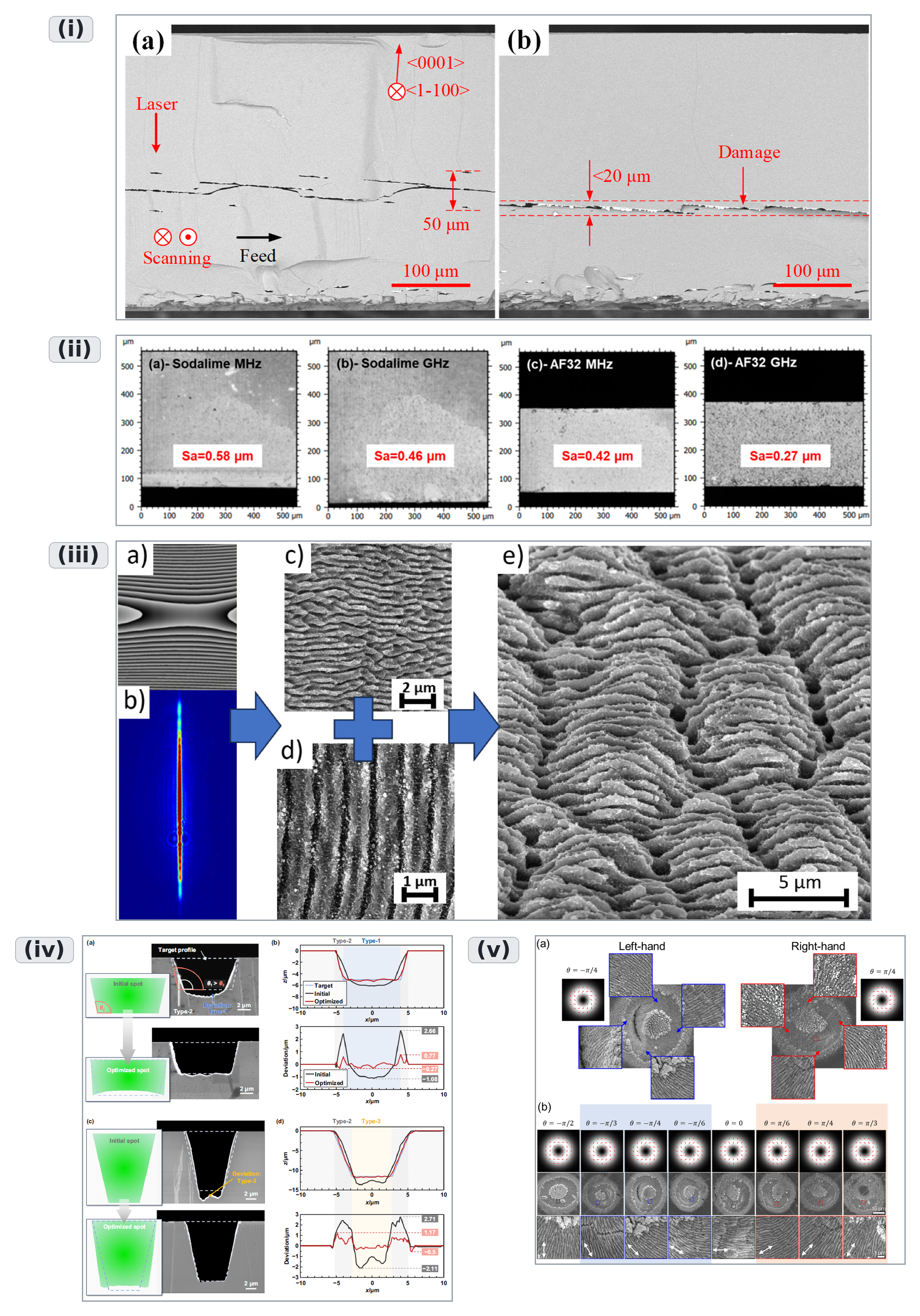}
  \caption{SLM-driven micromachining, from in-volume cutting to functional surface texturing. (i)~Stealth dicing in N-type SiC, modification confined below 20~\um. (ii)~GHz- versus MHz-burst Bessel cutting of glass. (iii)~Anti-adhesion LIPSS microgrooves in steel. (iv)~Adaptive-shaped microgrooves in SiC, sub-0.5~\um\ RMS error. (v)~Vector-vortex chiral LIPSS in tungsten. Panels (i)--(v) adapted from Wang \etal~\cite{wangIntegratedLCOSSLMBasedLaser2025}, Balage \etal~\cite{balageBesselBeamDielectrics2023}, Hauschwitz \etal~\cite{hauschwitzRapidLaserinducedNanostructuring2025}, Qiu \etal~\cite{qiuAdaptiveBeamshapingEnabled2025} and Kawaguchi \etal~\cite{kawaguchiFemtosecondVectorVortex2023}; \textcopyright{} 2025, 2023, 2025, 2025 and 2023 The Author(s), licensed under CC BY~4.0.}
  \label{fig:reuse_micromach}
\end{figure}

The toolkit now spends that uniformity on specific functional targets, with the target, not the modulator, setting the figure of merit. Hauschwitz \etal \cite{hauschwitzRapidLaserinducedNanostructuring2025} wrote a $500 \times 30$~\um\ top-hat line beam with 630~\nm\ laser-induced periodic surface structures (LIPSS) and 5~\um\ microgrooves, cutting \emph{Saccharomyces cerevisiae} adhesion by up to 99.88~\% (figure~\ref{fig:reuse_micromach}(iii)). Qiu \etal \cite{qiuAdaptiveBeamshapingEnabled2025} used adaptive shaping for 10~\um\ grooves in SiC at below 0.5~\um\ RMS profile error (figure~\ref{fig:reuse_micromach}(iv)), where fidelity rather than throughput is the metric. Kawaguchi \etal \cite{kawaguchiFemtosecondVectorVortex2023} wrote chiral vector-vortex LIPSS in tungsten, the annular vortex profile broadening the LIPSS-formation window \cite{bonseLaserinducedPeriodicSurface2017} (figure~\ref{fig:reuse_micromach}(v)). Three unrelated properties, adhesion, tolerance, and chirality, are each reached by reprogramming one SLM-CGH platform. This is where the cost-of-ownership inequality of section~\ref{sec:hardware_summary} (Eq.~\ref{eq:breakeven}) tips toward the programmable device. Every new target means another geometry change, and a fixed DOE would have to be re-fabricated to produce each one. These changes accumulate fastest where the texturing mix is varied, so that is where programmable optics pass break-even soonest.

Temporal shaping adds a further degree of freedom. Fang \etal \cite{fangPulseBurstGeneration2022} built a dual-SLM burst system giving a 323~\MHz\ intra-burst frequency with per-burst pattern selection. And Tan \etal \cite{tanThreedimensionalIsotropicMicrofabrication2023} used spatiotemporal focusing for three-dimensional isotropic microfabrication in photosensitive glass. Combined, these spatial and temporal degrees of freedom let texturing inherit the 100~\W\ industrial power budget established for SLM-driven parallel ablation \cite{hofmannDesignMultibeamOptics2020}. Its historic trade-off between throughput and flexibility has therefore visibly collapsed on the average-power axis.

\subsection{Photonic-device and waveguide writing}
\label{sec:micromach_waveguide}

Photonic-device writing carries the most stringent quality requirement of the four sub-classes \cite{wangFemtosecondLaserbasedProcessing2021, jiaRecentProgressFemtosecond2023}. A buried waveguide must support a target guided mode with sub-\um\ transverse resolution and a sub-dB~cm$^{-1}$ loss budget, in a single pass through hundreds of \um\ of dielectric and against spherical aberration at the interface. Here the binding constraint is aberration control at depth, so shaping buys resolution rather than speed. Surveys report shaping-enabled correction reaching features below the wavelength across material classes: 120~\nm\ in-plane and 500~\nm\ in depth in dielectric crystals \cite{jiaRecentProgressFemtosecond2023}, 100 and 200~\nm\ in two-photon polymerisation \cite{wangFemtosecondLaserbasedProcessing2021}, and sapphire-fibre gratings that operate to 1612~\degC \cite{zhaoReviewFemtosecondLaser2022}. Machining below the diffraction limit is now reported more widely \cite{huangSuperresolutionLaserMachining2025}.

Two demonstrations bracket the trade-off between speed and accuracy. He \etal \cite{heSlitBeamShaping2022} push accuracy hardest, using a slit-shaped beam to inscribe FBGs point by point, writing gratings of 1.07~\um\ pitch with more than 30~dB cladding-mode contrast at 0.3~dB insertion loss. Liu \etal \cite{liuFabricationSinglemodeCircular2021} sit nearer the speed end, writing 10~\um\ single-mode optofluidic channels at approximately 1~mm~s$^{-1}$. Further demonstrations extend the toolkit to single-pulse anisotropic lithography, fibre Rayleigh reflectors and compound-eye optics built from 3D spot arrays \cite{wangSinglepulseLithographyAmorphous2026, munkuevaFemtosecondLaserInscription2026, hayasakiVariableHolographicFemtosecond2005, sakakuraFabricationThreedimensional14Splitter2010, wangChalcogenideGlassIR2022, wangHolographicLaserFabrication2023, zhaiMicrofabricationBioinspiredCurved2020}, linking directly to TPP (section~\ref{sec:tpp}). A target for closed-loop correction during writing, if met, would decouple writing speed from the aberration penalty that builds up over successive passes \cite{cuiLightPeopleProfessor2022}.

\subsection{Parallel ablation (multi-spot CGH-driven)}
\label{sec:micromach_parallel}

Parallel ablation is the most direct expression of the trade-off between throughput and flexibility. An $N$-spot system runs at roughly $N$ times the single-spot rate, limited by the energy available per spot and the diffraction efficiency of the SLM. Those limits still leave the SLM the workhorse for industrial-scale ablation. Hofmann \etal \cite{hofmannDesignMultibeamOptics2020} set an early reference with an $8 \times 8 = 64$-spot array at 24--55~\um\ positioning error across a standard galvo scan field. Lutz \etal \cite{lutzEfficientUltrashortPulsed2021} then reached industrial-class operation. Twenty SLM spots at 1030~\nm\ and 100~\W\ removed material at 6~mm$^3$~min$^{-1}$, holding a near-constant efficiency of 0.06~mm$^3$~min$^{-1}$~\W$^{-1}$ across the 0.2--0.4~J~cm$^{-2}$ window. Together, these set the present baseline.

Jacob \cite{jacobPlanarLightValve2024} demonstrated an alternative modulator, a 1088-pixel planar light valve at 100~\kHz\ and approximately 1~mJ, giving 10~\um\ features at 2.5~\um\ edge accuracy and a greater than 30$\times$ throughput enhancement. Li \etal \cite{liHighQualityMicropatternPrinting2024} narrowed the per-spot quality trade-off with complex-amplitude holographic printing to 0.63~\um\ resolution. On the software side, machine-learning error compensation now learns the systematic phase distortions of the SLM and applies per-spot corrections at the pattern rate \cite{xuLightFieldModulation2023, buskeAdvancedBeamShaping2022, jacobDynamicBeamShaping2025}, closing the loop to the diffractive-network branch of section~\ref{sec:algo_dnn}. The same idea, matching the delivered energy to the feature being written, extends to keyhole stability in deep-penetration welding \cite{schmidtDynamicBeamShaping2024, bremerDesignImplementationDynamic2024}. Two open problems remain: deploying 300~\W-class SLMs into the industrial pipeline, where the best demonstrations sit at 100~\W, and integrating in-process monitoring with closed-loop pattern updates.

\subsection{Method transfer to adjacent processes}
\label{sec:micromach_transfer}

The same SLM/DOE/CGH toolkit carries into adjacent processes outside the strict ultrafast-micromachining scope, which confirms that the hardware and algorithms generalise. Laser marking fielded it first, a full decade before micromachining. Beck \etal \cite{beckAdaptiveLaserBeam2011} paired a $1024 \times 768$ SLM with an iterative Fourier-transform algorithm and closed-loop CCD feedback. This is the direct ancestor of today's camera-in-the-loop neural-holography methods \cite{pengNeuralHolographyCameraintheloop2020}, and later marking systems kept the same algorithmic spine \cite{lizotteBeamShapingMicro2003}. The toolkit has since moved into laser powder-bed fusion, where programmable shaping controls melt-pool geometry and microstructure \cite{biBeamShapingTechnology2023}. Grünewald \etal \cite{grunewaldGeneratingBricklikeMelt2024} formed a brick-like melt pool at 473~\W\ within 7~\% of target. And Esmaeilzadeh \etal \cite{esmaeilzadehArchitectedMicrostructuresUsing2025} shaped a profile in Ti-6Al-4V to steer a $\beta\to\alpha+\beta$ pathway and suppress martensite. Related studies comparing ring and Gaussian beams, 3~\kW\ green-laser copper welding and closed-loop defect control push the toolkit to macro-scale powers \cite{perez-ruizLaserBeamShaping2024, shiMicrostructuralControlMetal2020, mooreMicrostructureBasedModelingLaser2024, kaufmannTailoredLaserBeam2024, gunasegaramMachineLearningassistedInsitu2024}. Marking, powder-bed fusion, and welding all follow one pattern: a single programmable modulator with an algorithm to drive it, reshaping the process through the beam. These thermal processes lie outside the scope of this review, which is limited to athermal ultrafast micromachining. Nevertheless, they matter here only as evidence that the hardware and algorithms are general, not purpose-built for one process.

\newpage
\subsection{Synthesis}

Across the four sub-classes, programmable shaping produced four distinct gains. Glass cutting gained quality and depth. Laser texturing gained throughput. Photonic writing gained resolution. And parallel ablation gained combined high-$N$, high-power, and high-refresh operation. Wang \etal \cite{wangTracingFootprintsFemtosecond2026} trace the 2026 trajectory toward high-efficiency 2D/3D holographic femtosecond manufacturing. The practical frontier is now set by how well the modulator and its pattern-generation algorithm match the process, rather than by any single beam-shaping element.

\section{Beam shaping in two-photon polymerisation}
\label{sec:tpp}

Two-photon polymerisation (TPP) supplies the second regime in which the trade-off between throughput and flexibility of section~\ref{sec:challenges} breaks down. It is a key secondary application to the micromachining story of section~\ref{sec:micromach}, and an independent line of evidence for that collapse. TPP, also called two-photon lithography or multiphoton direct laser writing, is the additive counterpart to those subtractive processes \cite{maruoRecentProgressMultiphoton2008, farsariTwophotonFabrication2009, wangTwoPhotonPolymerizationLithography2023}. A femtosecond beam focused at high numerical aperture (NA) drives a two-photon-absorption reaction that polymerises a sub-\um\ voxel. Translating the focus in three dimensions then writes an arbitrary structure below the diffraction limit. Throughput therefore hinges on whether writing is serial or parallel, and the parallelisation depends directly on the modulator class and hologram algorithm of sections~\ref{sec:hardware} and~\ref{sec:algorithms} \cite{balenaRecentAdvancesHighSpeed2023, zhangHighThroughputTwoPhoton3D2024, kieferMultiphoton77focus2024, gu3DNanolithographyMetalens2025, wangFemtosecondLaserDirect2024}. Table~\ref{tab:tpp} consolidates the demonstrations discussed below.

\begin{table}[!htbp]
  \centering
  \footnotesize
  \setlength{\tabcolsep}{4pt}
  \caption{Two-photon-polymerisation demonstrations discussed in this section. Values are quoted from the cited work. A dash (--) marks an unstated quantity. The ``Throughput'' column reports each study's own figure in its original unit and is not comparable across rows. The \mbox{Feat./$\lambda$} column gives the smallest reported feature divided by the wavelength. It allows a rough comparison between studies, not an exact one.}
  \label{tab:tpp}
  \begin{tabularx}{\textwidth}{@{}l L C C L C c l@{}}
    \toprule
    \textbf{Study} &
    \textbf{Route} &
    \textbf{$\lambda$} &
    \textbf{Objective / NA} &
    \textbf{Parallelism} &
    \textbf{Throughput} &
    \textbf{Feat./$\lambda$} &
    \textbf{Feature size} \\
    \midrule
    Zhang \etal~\cite{zhangHighThroughputTwoPhoton3D2024} &
    Holographic multi-foci (LCoS-SLM) &
    1030~\nm &
    $60\times$ / NA~1.35 &
    $> 400$ foci &
    $1.49\times10^{8}$~voxel~s$^{-1}$ at 150~mm~s$^{-1}$ &
    0.44 &
    451~\nm \\
    Kiefer \etal~\cite{kieferMultiphoton77focus2024} &
    DOE + microlens array &
    790~\nm &
    $40\times$ / NA~1.4 &
    $7\times7=49$ foci, 19.5~mW/focus &
    $\sim10^{8}$~voxel~s$^{-1}$ at $\sim1$~m~s$^{-1}$ &
    0.87 &
    690~\nm\ (measured avg.\ voxel) \\
    Kim \etal~\cite{kimRapidPrintingNanoporous2023} &
    DMD projection &
    804~\nm, $\sim35$~\fs &
    $60\times$ / NA~1.25 &
    Single-exposure layer &
    $> 0.5$~mm$^2$~s$^{-1}$ &
    0.37 &
    $< 300$~\nm\ ($< 700$~\nm\ pores) \\
    Gu \etal~\cite{gu3DNanolithographyMetalens2025} &
    Metalens array (12~cm$^2$) &
    800~\nm &
    no objective; NA~1.0/0.8 (200/100~\um\ pitch) &
    $> 120{,}000$ foci &
    $> 5\times10^{7}$~parts~day$^{-1}$ &
    0.14 &
    113~\nm\ (262~\nm\ axial) \\
    Messer \etal~\cite{messerShoeboxsized3DLaser2024} &
    Two-step absorption (405~\nm\ CW) &
    405~\nm &
    $100\times$ / NA~1.4 &
    Single focus (serial), $< 1$~mW &
    $\sim1$~mm~s$^{-1}$ &
    0.25 &
    $\sim100$~\nm \\
    \bottomrule
  \end{tabularx}
\end{table}

\subsection{Parallel multi-spot writing}
\label{sec:tpp_multispot}

\begin{figure}[!h]
  \centering
  \includegraphics[width=0.98\textwidth]{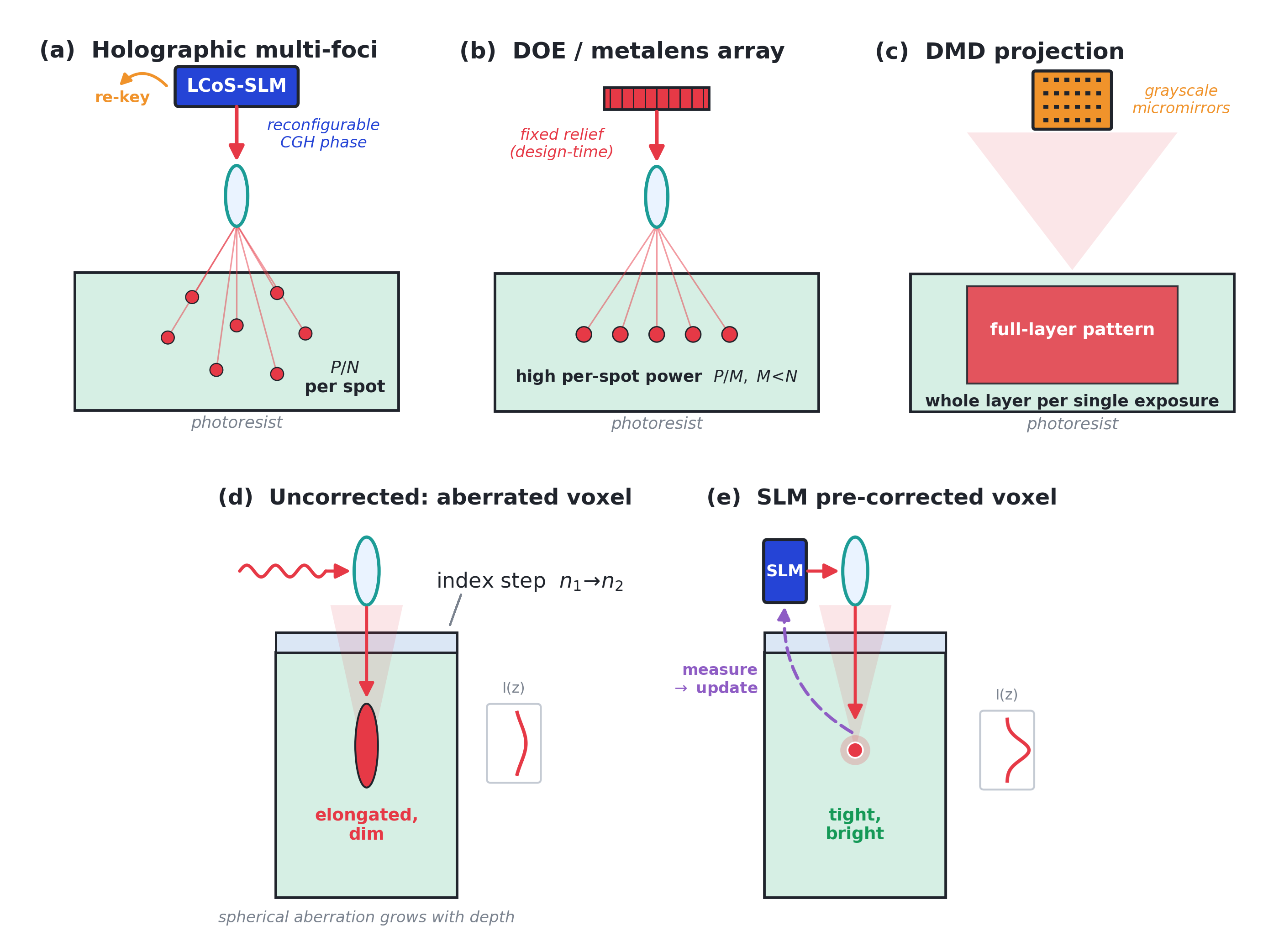}
  \caption{TPP process overview. (a--c) Three parallel-writing routes: reconfigurable holographic multi-foci on an LCoS-SLM, a fixed DOE- or metalens-defined focus array, and single-exposure DMD projection. (d--e) In-volume aberration correction: an uncorrected voxel elongated and dimmed by index mismatch, and the same voxel restored tight and bright by an iterative SLM pre-correction.}
  \label{fig:tpp_process}
\end{figure}

Throughput has always been the binding constraint on TPP. Single-focus serial scans at the \um~s$^{-1}$ speeds of the founding demonstrations build millimetre structures in hours and centimetre structures in days, which kept the technique in the prototype laboratory for two decades \cite{maruoRecentProgressMultiphoton2008, wangFlytrapInspiredPHDriven2022, liOpticalWaveguidesFabricated2023, juodkazisStereolithography3DMicrostructuring2002, wangTwoPhotonPolymerizationLithography2023}. Three parallel-writing routes now break that limit, and all three rest on programmable beam shaping (figure~\ref{fig:tpp_process}(a--c)).

The holographic route uses one CGH on an LCoS-SLM to place tens to hundreds of foci in the focal plane. The pattern stays reconfigurable from shot to shot, but this limits both the number of foci and the power each one carries \cite{kelemenParallelPhotopolymerisationComplex2007, vizsnyiczaiHolographicMultifocus3D2014, khoninaPerspectiveArtificialIntelligences2024}. Zhang \etal \cite{zhangHighThroughputTwoPhoton3D2024} set the current ceiling, writing at $1.49\times10^{8}$~voxels~s$^{-1}$ from more than 400 foci scanned at 150~mm~s$^{-1}$ with 451~\nm\ features. Interlayer crosstalk, the route's main weakness, has since been tackled by the 3D holographic algorithm of Wang \etal \cite{wang3DHolographicVolumetric2026}, which generates genuinely volumetric foci with above 90~\% reconstruction consistency in all spatial dimensions.

\begin{figure}[!htbp]
  \centering
  \includegraphics[width=\textwidth]{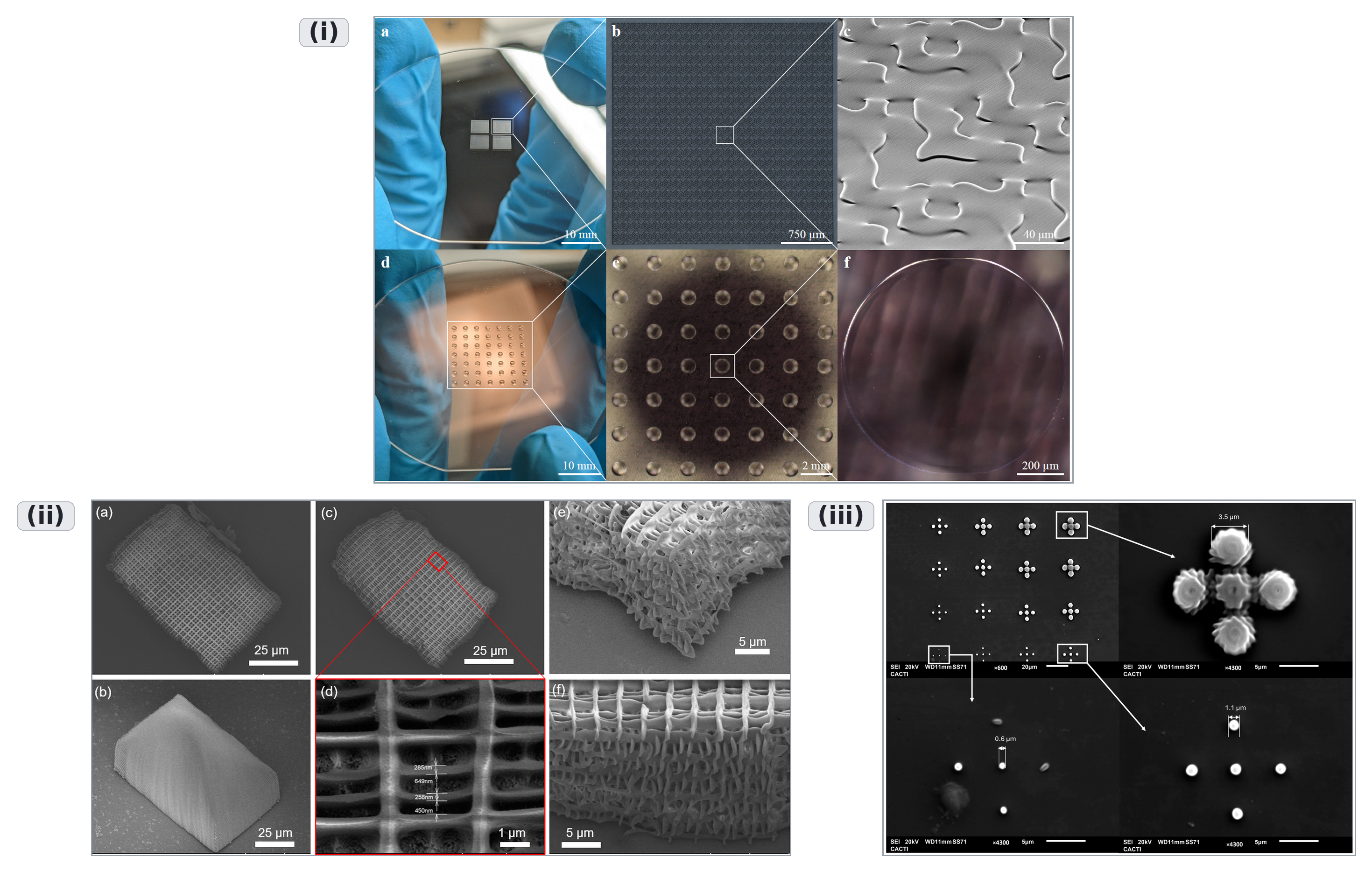}
  \caption{Demonstrations of parallel two-photon polymerisation. (i) A DOE-plus-microlens-array optic for $7\times7$ writing, with SEM close-ups of the written structures. (ii) DMD single-exposure projection of nanoporous woodpiles, with feature ($<300$~\nm) and pore ($<700$~\nm) sizes marked on the SEM images. (iii) SLM-based parallel writing of the focus, for higher productivity in realistic 2.5D and 3D workflows. Fixing the pattern buys higher per-spot energy (i) or whole-layer speed (ii). Panels (i)--(iii) adapted from Kiefer \etal~\cite{kieferMultiphoton77focus2024}, Kim \etal~\cite{kimRapidPrintingNanoporous2023} and Cuartero \etal~\cite{cuarteroEnhancingTwoPhotonPolymerization2026}; \textcopyright{} 2024, 2023 and 2026 The Author(s), licensed under CC BY~4.0.}
  \label{fig:reuse_tpp}
\end{figure}

The DOE-array route trades reconfigurability for a higher per-spot energy budget at a focus count fixed in the design. Kiefer \etal \cite{kieferMultiphoton77focus2024} generated a $7\times7$ array from a 3D-printed DOE and microlens array, reaching $\sim10^{8}$~voxels~s$^{-1}$ at $\sim1$~m~s$^{-1}$ with 19.5~mW per focus (figure~\ref{fig:reuse_tpp}(i)). The same fixed-pattern logic can also chase finer features. Xu \etal \cite{xuHighThroughputSubDiffractionLimitedTwoPhoton2026} combined ten-channel parallelism with peripheral photoinhibition and a hollow focus to narrow suspended nanowires while raising throughput over conventional direct writing.

 \newpage

The third route, single-exposure DMD projection, trades per-voxel selectivity for whole-layer speed. Kim \etal \cite{kimRapidPrintingNanoporous2023} printed a full layer per exposure at 804~\nm\ with $\sim35$~\fs\ pulses, exceeding 0.5~mm$^2$~s$^{-1}$ with sub-300~\nm\ features and sub-700~\nm\ pores in nanoporous 3D printing (figure~\ref{fig:reuse_tpp}(ii)). Kim and Saha's \cite{kimGrayscaleProjectionTwophoton2026} grayscale extension restores per-spot intensity control across more than 15,000 projected spots while keeping one pulse per layer, reporting 55~\nm\ nanowires at $1.7\times10^{9}$~voxels~s$^{-1}$.

The three routes sit at different points on the two-dimensional hardware space of section~\ref{sec:hardware}. The LCoS-SLM holds the high-flexibility, lower-power corner and can reprogram its focus pattern between adjacent layers \cite{xuLightFieldModulation2023}. The DOE and projection routes hold the high-throughput, fixed-pattern corner \cite{hofmannDesignMultibeamOptics2020}. Balena \etal's \cite{balenaRecentAdvancesHighSpeed2023}review adds that holographic multi-foci with optimised photo-initiators reach 100~mm~s$^{-1}$ at single-\um\ resolution, with STED-lithography and RAPID variants reaching 55~\nm\ and 40~\nm\ features at the same throughput \cite{wollhofen120NmResolution2013, malinauskasUltrafastLaserNanostructuring2013}. Cuartero \etal \cite{cuarteroEnhancingTwoPhotonPolymerization2026} confirm such gains from SLM-based parallelisation in realistic 2.5D/3D workflows (figure~\ref{fig:reuse_tpp}(iii)).

\subsection{Aberration correction in scattering and index-mismatched media}
\label{sec:tpp_aberration}

The second bottleneck is the aberration that the focus accumulates as it moves deeper into the resist (figure~\ref{fig:tpp_process}(d,e)). Refractive-index mismatch between immersion medium, cover glass, and resist adds spherical aberration that grows with depth and blurs the voxel \cite{jiaRecentProgressFemtosecond2023, qiaoFineOptimizationAberration2024}. The single-pass fix is a wavefront pre-correction on the same SLM that shapes the beam, set either from system geometry or, more powerfully, from a closed-loop measurement of the written voxel \cite{cuiLightPeopleProfessor2022, weinackerIterativePreCompensation3D2023, wangIntegratedLCOSSLMBasedLaser2025, danielLightFocusingScattering2019, galaktionovLaserBeamPropagation2015}. Weinacker \etal \cite{weinackerIterativePreCompensation3D2023} measured the deviation between the printed structure and the target with a confocal microscope at 200~\nm\ pitch, then subtracted it in the next iteration. This cut systematic geometry errors on 100~\um\ micro-lenses below the printing resolution itself \cite{maoEmergingFrontiersApplications2017}. The principle mirrors the crack pre-compensation of section~\ref{sec:micromach_glass} \cite{liaoDifferentiableDesignFreeform2024} and the camera-in-the-loop holography of section~\ref{sec:algo_citl} \cite{pengNeuralHolographyCameraintheloop2020, pengSpecklefreeHolographyPartially2021, yeomHighqualityPhaseonlyFourier2025}.

The wider transfer of adaptive optics from astronomy and microscopy into laser writing is traced by Booth in Cui \etal \cite{cuiLightPeopleProfessor2022}. Its key import is sensorless AO with an image-plane merit function, which drops the Shack--Hartmann sensor and has since reached in-process TPP correction in scattering media \cite{leiDoubledeformablemirrorAdaptiveOptics2012, rukosuevRealTimeCorrectionLaser2022}. A closed loop that updates the SLM at the \textmu s cadence of voxel-by-voxel writing does not yet exist, and section~\ref{sec:challenges} flags it as a near-term target \cite{wangReviewFemtosecondLaser2024, schmidtDynamicBeamShaping2024}. Liu \etal's \cite{liuModelDrivenDeepLearning2026} SMART HoloTPL takes the complementary route of embedding fabrication physics in the hologram generator, lowering speckle-driven non-uniformity before exposure and easing the in-process correction burden.

\subsection{Voxel and resolution control}

The third strand is the fundamental resolution limit. The diffraction-limited voxel scales as $\lambda/2{\rm NA}$ laterally and $\lambda/(2{\rm NA})^2$ axially, and the two-photon nonlinearity shrinks it by a further factor of about $\sqrt{2}$ in each direction. This places routine features in the 100--200~\nm\ regime for NA~1.4 immersion optics at 800~\nm \cite{maruoRecentProgressMultiphoton2008, wangTwoPhotonPolymerizationLithography2023}. Below 100~\nm, resolution is set less by the optics than by how hard the polymerisation threshold is pushed, and every sub-100~\nm\ route pays for it in dose control or photo-initiator chemistry. Nonlinear absorption alone reaches 80--120~\nm. Adding continuous-wave excitation to the pulse train reaches 40~\nm\ under tight dose control \cite{maoEmergingFrontiersApplications2017}, and radical quenching reaches a 60~\nm\ lateral feature against a typical 150~\nm\ floor \cite{zylaFrontiersLaserBased3D2024}. These finest features come only by narrowing the exposure window. Parallel writing (section~\ref{sec:tpp_multispot}) must instead spend that same latitude on keeping many foci uniform.

A separate route reaches the same resolution by photophysics rather than optics. Messer \etal \cite{messerShoeboxsized3DLaser2024} built a shoe-box-sized 3D nanoprinter around a 405~\nm\ continuous-wave GaN diode below 1~mW. An NA~1.4 objective focuses the beam and scans it at $\sim1$~mm~s$^{-1}$, reaching $\sim100$~\nm\ resolution without any femtosecond source. TPP-class resolution is therefore no longer locked to a femtosecond pulse, though not yet at holographic parallel throughput. Wang \etal's \cite{wangTwoPhotonPolymerizationLithography2023} review sets the baseline for the high-throughput demonstrations: sub-10~\nm\ roughness, $\sim1$~TW~cm$^{-2}$ peak intensity, and 0.5--5~nJ pulses.

\subsection{Arbitrary 3D beam shaping and the metasurface frontier}
\label{sec:tpp_metasurface}

The most striking recent step is the cross-over with the metasurface class. Gu \etal \cite{gu3DNanolithographyMetalens2025} built a 12~cm$^2$ metalens array that produces more than 120,000 cooperative foci over a centimetre-scale write field, with 262~\nm\ axial linewidths and 113~\nm\ features. An upstream LCoS-SLM selects the active subset of foci on each layer. In an industrial demonstration, the system replicated more than 50 million microparticles per day, beating every previous single-objective platform. It stacks a reconfigurable CGH in series with fixed metasurface optics, dissolving the old split between fixed patterns for throughput and SLM writing for flexibility \cite{zhangHighThroughputTwoPhoton3D2024, kieferMultiphoton77focus2024}.

SLM-programmed 3D spot arrays also fabricate functional photonic devices directly. Wang \etal \cite{wangHolographicLaserFabrication2023} wrote a 61-lenslet concave compound-eye array at 185~nJ per spot with sub-96~\nm\ surface roughness. This is the strongest single-run functional optic among SLM-driven 3D-array writes. The cross-over with photonic-device writing (section~\ref{sec:micromach_waveguide}) runs through aberration correction at depth. It lets the systems reviewed by Jia and Chen \cite{jiaRecentProgressFemtosecond2023} and the FBG inscription of He \etal \cite{heSlitBeamShaping2022} reach features below the wavelength well inside the material. That same correction also sets a trade-off between precision and speed. He \etal\ \cite{heSlitBeamShaping2022} work at the precision end, while the optofluidic waveguides of Liu \etal \cite{liuFabricationSinglemodeCircular2021} sit at the speed end.

Applied TPP has matured alongside the throughput and resolution work. One demonstration is a flytrap-inspired hydrogel actuator, driven by pH, with a 230~\nm\ minimum linewidth and 1.2~s response \cite{wangFlytrapInspiredPHDriven2022}. Another writes chiral lattices with an SLM vortex beam, reaching 66~\% helical dichroism for detecting orbital angular momentum \cite{liChiralLithographyVortex2024}. A third produces LiNbO$_3$ waveguides below 0.5~dB~cm$^{-1}$ loss \cite{liOpticalWaveguidesFabricated2023, floreaNewFrontiersMaterials2023}.

The remaining hardware constraint is the damage threshold of the static optics that follow the SLM. Wang \etal \cite{wangVortexfieldEnhancementHighthreshold2024} report a geometric metasurface above 99.4~\% transmittance with a 68.0~J~cm$^{-2}$ LIDT at 1064~\nm\ under 6~ns pulses, so this component is not the per-pulse bottleneck under nanosecond loading. That result does not transfer to the femtosecond-burst regime of ultrafast micromachining. Whether the metasurface survives \MHz-burst fs illumination at at the throughput of Gu's metalens array \cite{gu3DNanolithographyMetalens2025} is an open question the present literature does not resolve \cite{crottiGiantUltrafastDichroism2024, lengMetasurfaceMirrorsBased2024, lafargueInVolumeGlassModification2024}.

\subsection{Synthesis}

Across the four strands, TPP over the last year has raised throughput, voxel resolution, and in-volume aberration tolerance together. These three metrics once traded against each other. The pattern mirrors the four subsections of section~\ref{sec:micromach}, and it is the second-strongest evidence that the frontier between throughput and flexibility identified in section~\ref{sec:intro} has collapsed. Three gaps remain. The first is aberration update at MHz rates, fast enough to close the simulation-to-experiment loop for features below 100~\nm\ at industrial scan speeds \cite{chengFlexibleTunedMultifocus2024}. The second is measuring the damage threshold of the static optics placed after the SLM, metasurfaces included, under femtosecond-burst illumination at the throughput of the Gu metalens array \cite{gu3DNanolithographyMetalens2025}. The third is reproducible operation over long runs at the energy per spot that holographic writing with more than 100 foci demands \cite{lutzEfficientUltrashortPulsed2021, zuoHighPerformanceNIRLaserBeam2025, tangExtendingOperationalLimit2025}.

\section{Hybrid and computational architectures}
\label{sec:hybrid}
\begin{figure}[!h]
  \centering
  \includegraphics[width=0.9\textwidth]{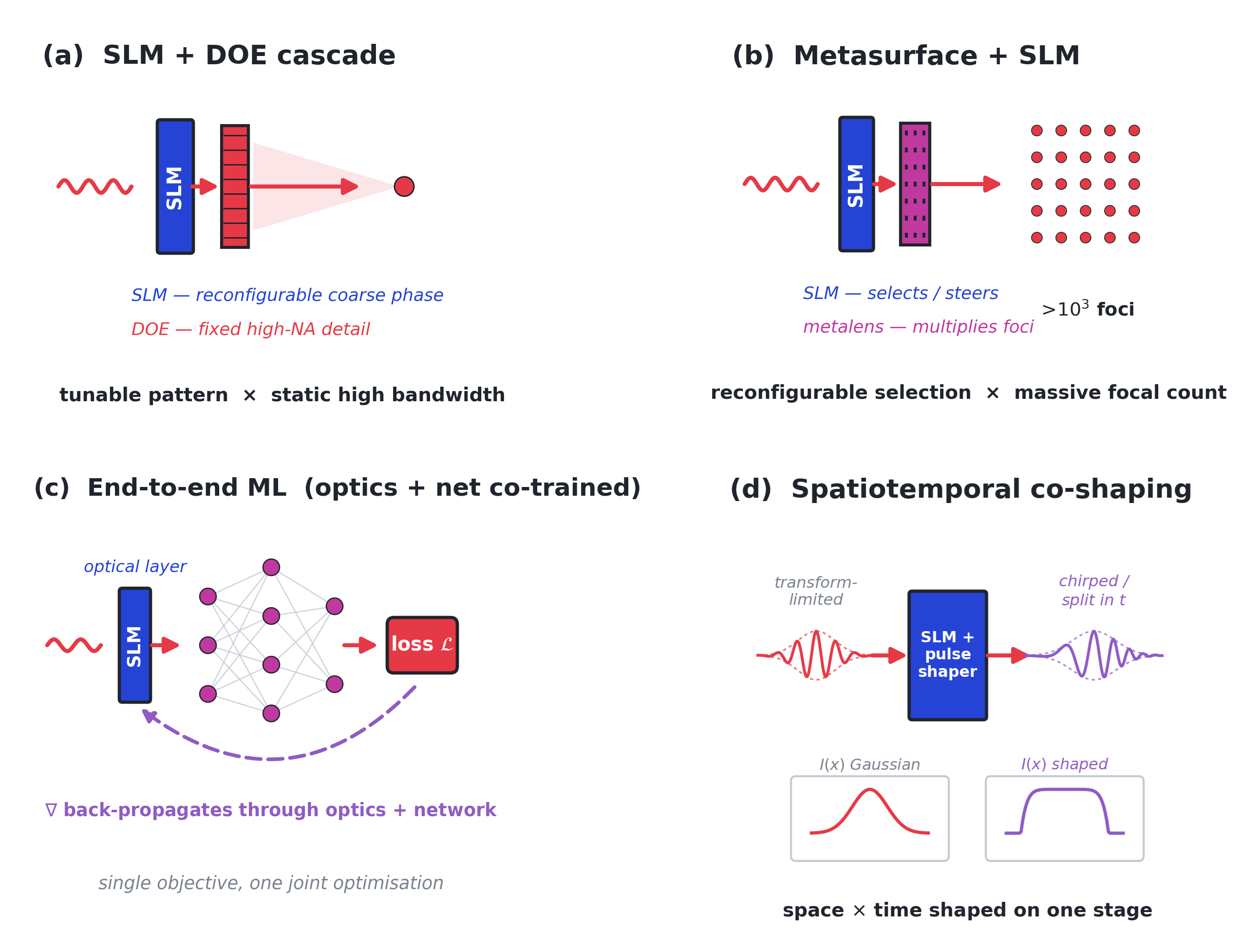}
  \caption{The four hybrid beam-shaping architectures. Each stacks two or more layers so that a later layer relaxes a constraint of the one before it, reaching a throughput and flexibility neither layer holds alone.}
  \label{fig:hybrid_stack}
\end{figure}

Sections~\ref{sec:hardware} to~\ref{sec:tpp} treated the hardware classes, the algorithms, and the applications as if each could be optimised alone. The strongest recent results come instead from architectures that stack two or more of these layers and tune them together. These stacks are the principal evidence for the collapse of the frontier between throughput and flexibility identified in section~\ref{sec:intro}. This chapter covers four of them: SLM\,+\,DOE cascades, metasurface\,+\,SLM stacks, end-to-end machine-learning systems, and spatiotemporally co-shaped pulse trains. In each, a later optical layer relaxes a constraint that limits the layer before it, so the stack reaches an operating point that neither layer holds alone (figure~\ref{fig:hybrid_stack}) \cite{dunDynamicLaserBeam2018, buskeAdvancedBeamShaping2022}.

One system-level route reaches the same corner without stacking devices. Coherent beam combining shapes an output by controlling the relative phase of many emitters rather than modulating a single wavefront. Weber \etal \cite{weberBasicPropertiesHighDynamic2025} shaped a 6--14~kW combined output by switching the phase between fibre channels at 80~MHz, re-addressing the beam on a ${\sim}12.5$~ns time constant. Additionally, Adamov \etal \cite{adamovLaserBeamShaping2021} showed the same amplitude--phase principle on a lower-power array. Cavity-internal shaping with an external pump SLM has similarly excited over 100 Hermite--Gaussian modes up to HG$_{25,27}$ \cite{schepersSelectiveHermiteGaussian2019, burgerImplementationSpatialLight2014, shenWavelengthtunableHermiteGaussianModes2018}. Shaping at the system level, coherent combining complements the LCoS-SLM and DM rather than replacing them. It is the only route yet to reach the high-refresh, high-power corner that the device stacks approach from below, and its strongest micromachining case is parallel-spot ablation at industrial average power (section~\ref{sec:micromach_parallel}) \cite{hofmannDesignMultibeamOptics2020, lutzEfficientUltrashortPulsed2021}.

\newpage

\subsection{SLM\,+\,DOE cascades}

The foundational hybrid pairs a programmable SLM with a static DOE. The SLM supplies per-shot reconfigurability, while the DOE supplies a high-bandwidth pre-shaped field the SLM need not synthesise from scratch. Dun \etal \cite{dunDynamicLaserBeam2018} showed the principle with hybrid holograms on a single SLM, reaching a 0.5~mm reconstruction range while keeping dynamic control over a statically encoded grating. The cascade has since moved into industrial micromachining, where the DOE fixes the multi-spot base pattern, and the SLM performs per-spot fine adjustment \cite{hofmannDesignMultibeamOptics2020, lutzEfficientUltrashortPulsed2021, schmidtDynamicBeamShaping2024, lafargueInVolumeGlassModification2024}.

Kiefer \etal \cite{kieferMultiphoton77focus2024} applied the DOE\,+\,microlens-array variant to TPP. A printed DOE generates a $7 \times 7$ focus grid, a microlens array images each focus into resist, and an upstream SLM sets the per-frame illumination. Read against the all-holographic route of Zhang \etal \cite{zhangHighThroughputTwoPhoton3D2024}, the point is that the DOE tolerates more power than an SLM and so carries a higher per-focus budget. That is why the cascade reaches a throughput neither layer delivers alone \cite{buskeAdvancedBeamShaping2022}.

\subsection{Metasurface\,+\,SLM stacks}

The clearest metasurface\,+\,SLM hybrid is the Gu \etal \cite{gu3DNanolithographyMetalens2025} metalens-array platform (section~\ref{sec:tpp_metasurface}). An upstream LCoS-SLM is stacked in front of a 12~cm$^{2}$ static metalens array. The SLM controls which of those foci are illuminated, updating the selection for each TPP layer at its own refresh rate. The system therefore inherits the SLM's layer-by-layer reconfigurability and the metasurface's dense, high-fidelity focal generation at once. It is the single clearest case of the active--passive binary dissolving inside one stack \cite{gu3DNanolithographyMetalens2025}. \newpage

\begin{figure}[!h]
  \centering
  \includegraphics[width=0.82\textwidth]{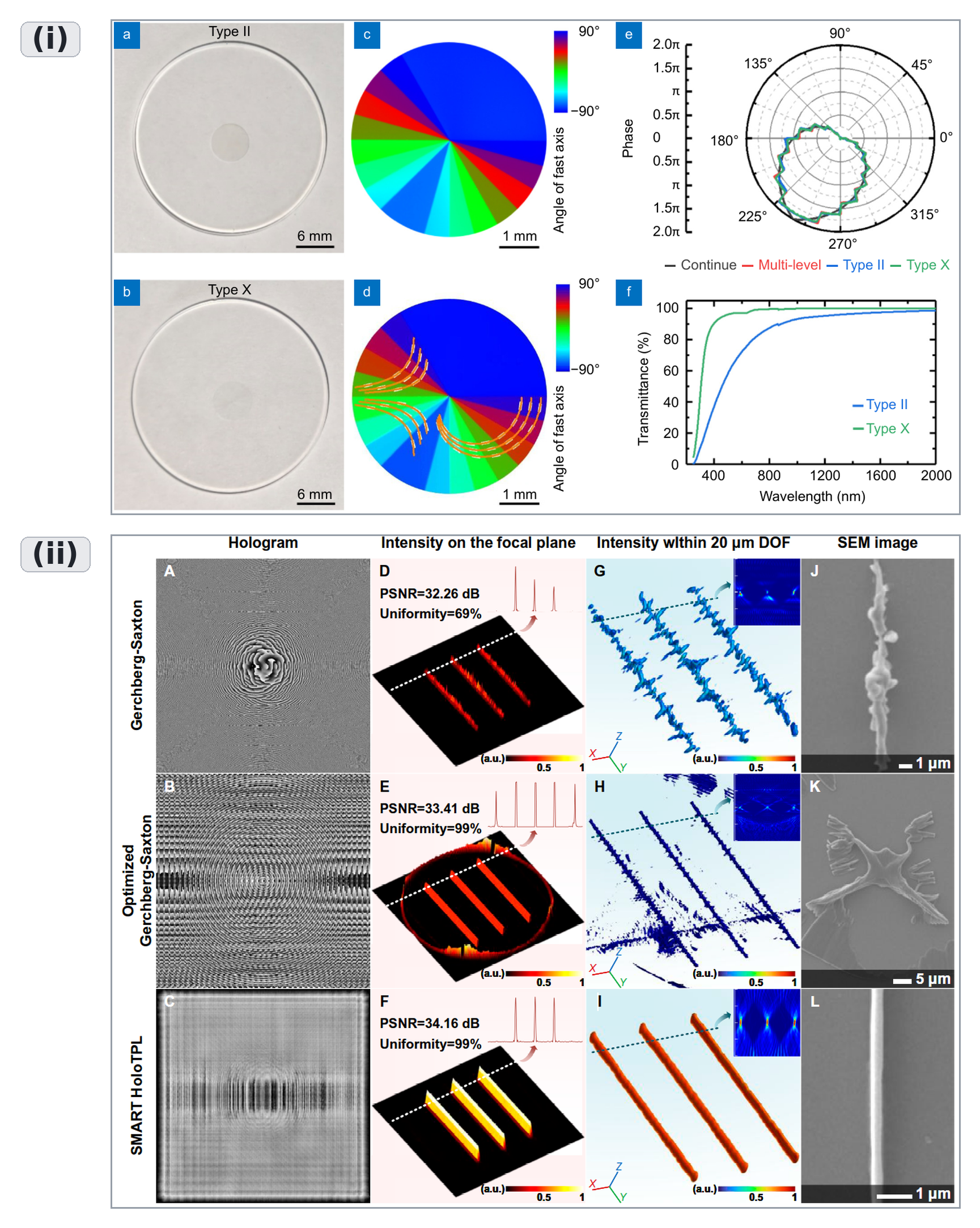}
  \caption{Two hybrid architectures from recent work. (i)~A geometric metasurface as the static shaping layer in a metasurface\,+\,SLM stack: two 6~mm silica disks with their fast-axis maps and greater than 99~\% near-infrared transmission. (ii)~End-to-end model-driven learning (SMART HoloTPL): 99~\% uniformity and speckle-free writing, against the pronounced speckle of plain Gerchberg--Saxton. Panels (i) and (ii) adapted from Wang \etal~\cite{wangVortexfieldEnhancementHighthreshold2024} and Liu \etal~\cite{liuModelDrivenDeepLearning2026}; \textcopyright{} 2024 and 2026 The Author(s), licensed under CC BY~4.0.}
  \label{fig:reuse_hybrid}
\end{figure}

A second variant uses the metasurface as a static wavelength multiplexer with the SLM supplying the per-channel hologram. The geometric metasurface of Wang \etal \cite{wangVortexfieldEnhancementHighthreshold2024} (figure~\ref{fig:reuse_hybrid}(i)) tolerates high fluence, showing that metasurfaces placed after the SLM are no longer the damage bottleneck under nanosecond bursts. The active-metasurface class points further. Crotti \etal \cite{crottiGiantUltrafastDichroism2024} induced a transient $\pi/2$ phase shift at 180~\textmu J~cm$^{-2}$ on a 2~\ps\ timescale. The metasurface is itself reconfigurable, and in that case the SLM may act as a control laser rather than the principal shaper \cite{khoninaPerspectiveArtificialIntelligences2024, lengMetasurfaceMirrorsBased2024}.

 \newpage

Fan \etal \cite{fanSpatialLightModulator2026} inverted the arrangement recently. A downstream DMD projects a time-varying pattern $O(x,y,\lambda,t)$ onto a static prediscretised TiO$_2$ metasurface carrying $M(x,y,\lambda)$, giving an output $R = O \cdot M$. Modulation fidelity follows the 756~nm metasurface pitch rather than the $\geq 5$~\textmu m DMD pixel. This optically addressed metasurface SLM reaches a spatiotemporal product density of $2.3 \times 10^{12}$~pixels~s$^{-1}$~cm$^{-2}$ while running at ${\sim}13$~kHz. That is three orders above the best commercial LCoS entry, and above the ${\sim}10^{12}$ threshold set for real-time holographic display. This resets the pixel axis of the SLM frontier. The limit is no longer the thickness of the LC layer or the MEMS hinge, but the pitch, on the scale of the wavelength, that a metasurface provides routinely \cite{fanSpatialLightModulator2026}.

At the opposite speed extreme, Hail, Michaeli and Atwater \cite{hailUltrafastReconfigurableAlloptical2026} demonstrated all-optical modulation on a high-$Q$ Mie-resonant metasurface driven by a 100-\fs, 10-\kHz\ amplifier. The mechanism is the instantaneous optical Kerr effect, which removes the picosecond carrier-relaxation time that limits semiconductor SLMs. This gives a 74-\fs\ response and $\pm 13^{\circ}$ steering, roughly ten orders faster than LC director relaxation. A patterned pump imposes an arbitrary 2D phase grating, so the effective refresh rate equals the drive laser's repetition rate. The device serves ultrafast-pulse shaping and single-shot spatiotemporal control rather than CW industrial processing.

\subsection{End-to-end machine-learning systems}

The third class is fully computational: the optical and digital layers are trained as one. Buske \etal \cite{buskeAdvancedBeamShaping2022} trained a diffractive neural network whose hidden layers are programmable phase masks at $1800 \times 1800$ neurons per axis. Liao \etal \cite{liaoDifferentiableDesignFreeform2024} made the DOE phase profile differentiable against a downstream simulation under a B-spline smoothness constraint. Rahman and Özcan \cite{rahmanIntegrationProgrammableDiffraction2024} generalised this to a diffractive language that jointly optimises programmable diffraction and digital networks across coherent, partially coherent and incoherent illumination \cite{guoVectorialDigitelligentOptics2024}.

The wider AI-photonics literature supplies context. Mahmoud \etal \cite{mahmoudAIdrivenPhotonicsUnleashing2024} report a physics-informed network for modal analysis at 0.69~\% relative error, together with inverse design of grating couplers by reinforcement learning. Manufacturing reviews flag real-time monitoring, defect prediction, and parameter optimisation as the most promising targets \cite{murzinArtificialIntelligenceDrivenInnovations2024, murzinComputerScienceIntegrations2024, murzinDigitalEngineeringPhotonics2024}. Jacob \etal \cite{jacobDynamicBeamShaping2025} demonstrated a diffractive network that adapts across 915, 1064 and 1550~\nm. It generalises across wavelengths in a way earlier machine-learning CGH approaches could not \cite{khoninaPerspectiveArtificialIntelligences2024}. The clearest transfer into direct writing is the SMART HoloTPL method of Liu \etal \cite{liuModelDrivenDeepLearning2026} (section~\ref{sec:tpp_aberration}). Its loss function embeds the polymerisation physics, reaching 99~\% uniformity and writing free of speckle where a plain Gerchberg--Saxton reconstruction leaves pronounced speckle (figure~\ref{fig:reuse_hybrid}(ii)).

\subsection{Spatiotemporally co-shaped pulse trains}
\label{sec:hybrid_spatiotemporal}

The fourth hybrid optimises space and time together rather than two spatial-shaping layers. Its extra degree of freedom lets the intensity distribution vary during the pulse. Pulse-front tilt and spatiotemporal focusing (STF) couple the spatial and spectral coordinates so the frequency components overlap only near the geometric focus. Axial confinement is then set mainly by temporal recompression and only weakly by numerical aperture. This decouples axial from lateral confinement, so each can be tuned almost independently. At the device level, Panuski \etal \cite{panuskiFullDegreeoffreedomSpatiotemporal2022} demonstrated a photonic-crystal cavity array near the joint space--bandwidth and time--bandwidth limits. It switches in nanoseconds at femtojoule energies across 64 resonators, pointing toward space--time shaping in a single stage \cite{mounaixSpatiotemporalCoherentControl2015}.

\begin{figure}[!htbp]
  \centering
  \includegraphics[width=0.8\textwidth]{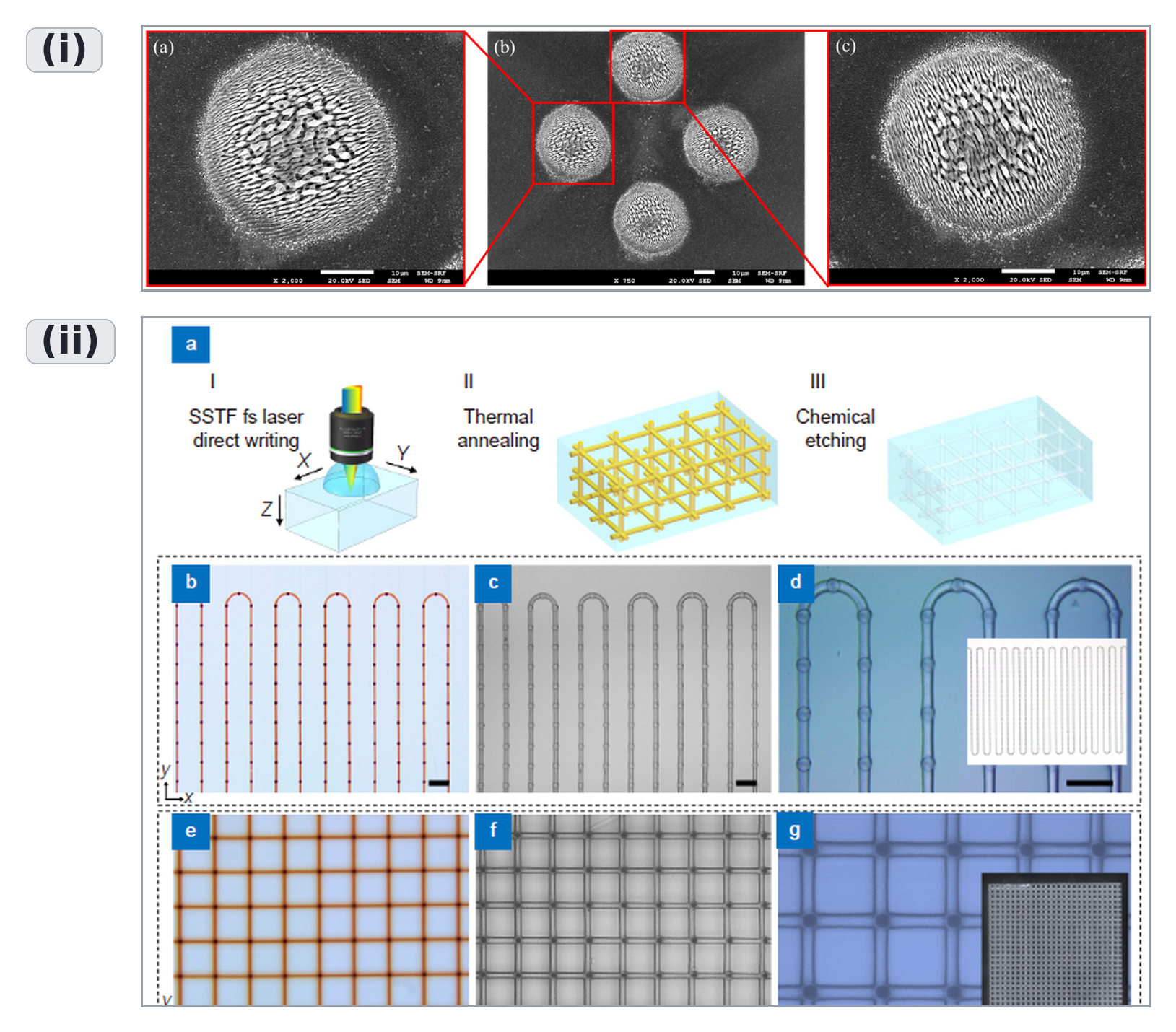}
  \caption{Spatiotemporally co-shaped processing. (i)~LIPSS written simultaneously with orthogonal polarisations on stainless steel on a two-SLM pulse-burst platform, shown at three magnifications. (ii)~Spatiotemporally focused \fs\ writing of almost isotropic 3D microchannels in glass, with a process schematic and micrographs of the embedded channels. Panels (i) and (ii) adapted from Fang \etal~\cite{fangPulseBurstGeneration2022} and Tan \etal~\cite{tanThreedimensionalIsotropicMicrofabrication2023}; \textcopyright{} 2022 and 2023 The Author(s), licensed under CC BY~4.0.}
  \label{fig:reuse_spatiotemporal}
\end{figure}

Tan \etal \cite{tanThreedimensionalIsotropicMicrofabrication2023} used STF of \fs\ pulses at a high repetition rate to write almost isotropic 3D features in glass, tuning resolution between 8 and 22~\um\ at 200~\um~s$^{-1}$. The isotropic voxel follows directly from the decoupling of axial and lateral confinement described above (figure~\ref{fig:reuse_spatiotemporal}(ii)). Schmidt \etal \cite{schmidtDynamicBeamShaping2024} made the broader argument. The coupling between transient energy input and the addressed process feature is the dominant degree of freedom in materials processing, so spatial and temporal shaping cannot be separated. Fang \etal \cite{fangPulseBurstGeneration2022} showed the engineering corollary. On a platform with two SLMs, they converted a 5~\kHz\ pulse train into a 323~\MHz\ burst and shaped each pulse in the burst independently. Space and time are therefore shaped together on one reconfigurable stage (figure~\ref{fig:reuse_spatiotemporal}(i)) \cite{lutzEfficientUltrashortPulsed2021, lafargueInVolumeGlassModification2024}.

\subsection{Synthesis}

The four architectures are together the principal evidence for the trade-off collapse argued in section~\ref{sec:intro}, and none is visible from a binary active-versus-passive view of beam shaping. Each reaches an operating point superior to either component alone. SLM\,+\,DOE keeps SLM reconfigurability at DOE throughput \cite{dunDynamicLaserBeam2018, kieferMultiphoton77focus2024}. Metasurface\,+\,SLM scales the focus count by orders of magnitude while keeping per-layer pattern selection \cite{gu3DNanolithographyMetalens2025}. End-to-end ML jointly designs the optical and digital layers \cite{buskeAdvancedBeamShaping2022, liaoDifferentiableDesignFreeform2024, rahmanIntegrationProgrammableDiffraction2024}. And spatiotemporal co-shaping reaches a dimension the spatial toolkit alone cannot \cite{tanThreedimensionalIsotropicMicrofabrication2023, schmidtDynamicBeamShaping2024, fangPulseBurstGeneration2022}. Han \etal \cite{hanFemtosecondLaserNonDiffractingBeam2026} close the loop further, using phase-modulated quasi-Bessel \fs\ lithography with a depth of focus more than ten times that of a focused Gaussian. The toolkit that shapes the beam and the process that manufactures the next reconfigurable element thus begin to merge. Every architecture at the current frontier is a stack, and each earns its performance by letting a later layer relax a constraint of the one before it. It is in these stacks, and not in any single device, that the active--passive binary of section~\ref{sec:intro} has most visibly dissolved, and against them that the axis-by-axis verdict of section~\ref{sec:challenges} is set.

\section{Challenges and outlook}
\label{sec:challenges}

The collapse of the historic trade-off is real, but it is partial. Table~\ref{tab:master_verdict} records the verdict for each application regime. It is confined to one axis and two regimes. On the average-power axis, in industrial parallel ablation and in high-throughput two-photon polymerisation, a programmable device now reaches a throughput once reserved for static optics. This shows that the trade-off between throughput and flexibility of section~\ref{sec:intro} is breaking down. Everywhere else the trade-off persists, and on the peak-fluence axis that ultimately governs micromachining, none has been shown at all. Economics narrows the claim even where it holds. The break-even model of section~\ref{sec:hardware_summary} (Eq.~\ref{eq:breakeven}) still favours a static optic wherever the geometry is fixed, and the volume is high. This breakdown therefore widens the range of conditions in which programmable optics are the better choice, but it does not displace static optics wholesale.

Two regimes account for the shift. In industrial parallel ablation at the 100~\W\ class, SLM holograms split the beam into 20 spots and remove material at 6~mm$^3$~min$^{-1}$, holding a near-constant efficiency per watt \cite{lutzEfficientUltrashortPulsed2021}. The modulator hardware has since been pushed further, to 210~\W\ under pulsed loading \cite{tangExtendingOperationalLimit2025}, and in continuous-wave power-handling tests to $\approx 380$~\W\ \cite{zuoHighPerformanceNIRLaserBeam2025} and beyond a kilowatt \cite{wolenskiKilowattAveragePower2025}. In high-throughput parallel two-photon polymerisation, holographic-multi-foci and DOE-array routes have each reached of order $10^{8}$~voxels~s$^{-1}$, and a metalens-array platform reports a further order of magnitude \cite{zhangHighThroughputTwoPhoton3D2024, kieferMultiphoton77focus2024,gu3DNanolithographyMetalens2025}. In both, the decisive lever was the co-design of the modulator with its pattern-generation algorithm rather than either taken alone. Two of the strongest figures, the 1.4~\kW\ SLM and the 120{,}000-focus platform, rest on a single group and await independent replication \cite{wolenskiKilowattAveragePower2025, gu3DNanolithographyMetalens2025}. Even with both set aside, the verdicts still hold, resting on Tang's \cite{tangExtendingOperationalLimit2025} 210~\W\ demonstration and on the routes of Zhang \cite{zhangHighThroughputTwoPhoton3D2024} and Kiefer \cite{kieferMultiphoton77focus2024}. The trade-off persists in sub-micrometre photonic-device writing, in the long-duration reliability of high-power SLMs, and in the fs-burst damage threshold of the static optics downstream of the modulator \cite{jiaRecentProgressFemtosecond2023, weinackerIterativePreCompensation3D2023, buskeAdvancedBeamShaping2022}.

Co-design has a hard limit that no hardware or algorithm can lift. Wherever the binding constraint is the material's own response time, faster reprogramming only redistributes a fixed budget. Three such ceilings are already visible. In high-repetition-rate processing, heat accumulation caps the usable pulse rate once the inter-pulse interval falls below the substrate's thermal-diffusion time \cite{shinReviewHighprecisionFemtosecond2024, raciukaitisUltraShortPulseLasers2021}. Weber \etal \cite{weberDeterministicControlDynamic2026} make this quantitative, tabulating characteristic process frequencies from \SI{1}{\kilo\hertz} to \SI{10}{\mega\hertz} for aluminium, iron and titanium, above which a shaping degree of freedom acts only through its time-averaged profile. In two-photon polymerisation, resist depletion and re-initiation kinetics bound the voxel dwell rate, which is why the fastest demonstrations spend their gains on more parallel foci \cite{wangTwoPhotonPolymerizationLithography2023}. In dielectric ablation, plasma shielding of the pulse tail sets a fluence-coupling ceiling independent of beam shape \cite{raciukaitisUltraShortPulseLasers2021}.

The residual gaps fall on the same seven axes on which table~\ref{tab:six_axis_master} scores the hardware classes, so each verdict below reads directly against the correspondingly named specification row.

\textbf{Reconfigurability.} The bottleneck has moved from hardware to algorithm. Modulators support kilohertz refresh, but iterative Gerchberg--Saxton takes seconds and learned inference trades run-time for training-time \cite{whyteExperimentalDemonstrationHolographic2005, wuAdaptiveWeightedGerchbergSaxton2021, yuUseDeepLearning2025}.

\textbf{Average power and peak fluence.} The ceiling under continuous-wave illumination now reaches the kilowatt scale. Peak-fluence behaviour under femtosecond bursts at MHz repetition rates, the regime that governs micromachining, remains uncharacterised for every programmable class \cite{wolenskiKilowattAveragePower2025, lafargueInVolumeGlassModification2024}.

\textbf{Optical efficiency.} The SLM at $0.98$ now matches the deformable mirror above $0.99$ and multi-level DOEs above $0.95$. Metasurfaces still lag, at $0.66--0.92$ for single-function designs and lower once one element multiplexes several channels \cite{bremerDesignImplementationDynamic2024, gu3DNanolithographyMetalens2025,chenLightweightEfficientBeamshaping2025}.

\textbf{Beam-quality fidelity.} Camera-in-the-loop learned holography matches iterative-GS quality at far shorter inference time, though cross-system transferability of a trained model is incomplete \cite{pengNeuralHolographyCameraintheloop2020, pengSpecklefreeHolographyPartially2021, yeomHighqualityPhaseonlyFourier2025}.
\newpage
\textbf{Three-dimensional programmability.} A hybrid of a metasurface and an SLM reaches a write field of centimetre scale, but only with a fixed metasurface design. No fully reconfigurable three-dimensional shaper yet works over a field that large \cite{gu3DNanolithographyMetalens2025, crottiGiantUltrafastDichroism2024, liaoDifferentiableDesignFreeform2024}.

\textbf{Capital expenditure and integration.} The kilowatt SLM is a specialised, water-cooled assembly, while commodity hardware stays capped at tens of watts. Scaling the demonstrated device to routine deployment is a problem of engineering and investment, not of physics \cite{wolenskiKilowattAveragePower2025, buskeHighFidelityLaser2023}.

\textbf{Per-pattern computational cost.} This is the one axis where a static optic is strictly superior, and the gap has not closed. A fixed DOE computes nothing per part, while every programmable device pays a recurring cost each time it is reprogrammed. For the learned CGH pipelines that drive this shift, that cost has barely been benchmarked against the iterative GS baseline they replace \cite{yuUseDeepLearning2025, khoninaPerspectiveArtificialIntelligences2024}.

The peak-fluence axis carries the single largest caveat, because its damage is mechanism-specific and none of the reported figures is the quantity that governs micromachining. In the LCoS-SLM, the failure is a thermally driven disruption of the nematic layer. Single-shot femtosecond damage sets in above $1$~J\,cm$^{-2}$ at 1030~\nm\ \cite{ramousseFemtosecondDamageThreshold2021}, while a thermo-optically addressed architecture reaches $500$~GW\,cm$^{-2}$ \cite{ramousseThermoopticalSLMDamageThreshold2025}. In the DMD, it is hinge and coating ablation, characterised only at picosecond duration \cite{schwarzImpactThresholdAssessment2021}. In the metasurface, it is meta-atom ablation, known only from a nanosecond LIDT \cite{wangVortexfieldEnhancementHighthreshold2024}. The figure that actually matters is the fluence a cooled device can survive under femtosecond bursts at MHz repetition rates, sustained over a full duty cycle. No high-power demonstration reports it. A standardised protocol for femtosecond-burst LIDT would close this gap. It would drive a rated device with a representative burst train, ramp the peak fluence in fixed steps at constant average power, and monitor the optic in situ over hours rather than seconds, with survival defined as no irreversible change. Until such a protocol exists, the peak-fluence axis rests on inference from average-power proxies, and this review therefore scores it open.

{\footnotesize
\setlength{\tabcolsep}{4pt}
\begin{longtable}{@{}
  >{\raggedright\arraybackslash}p{0.148\textwidth}
  >{\raggedright\arraybackslash}p{0.102\textwidth}
  >{\raggedright\arraybackslash}p{0.158\textwidth}
  >{\raggedright\arraybackslash}p{0.325\textwidth}
  >{\raggedright\arraybackslash}p{0.195\textwidth}
@{}}
  \caption{Consolidated verdict on the trade-off between throughput and flexibility, by regime. ``Collapsed'' = a programmable device now matches a throughput or quality once reserved for static optics. ``Persists'' = the historic coupling still binds. ``Status quo ante'' summarises the pre-programmable baseline from the corresponding section text (sections~\ref{sec:micromach}--\ref{sec:tpp}).}
  \label{tab:master_verdict}\\
    \toprule
    \textbf{Item} & \textbf{Verdict} & \textbf{Status quo ante} & \textbf{Status 2026 / decisive evidence} & \textbf{What still blocks it} \\
    \midrule
    \endfirsthead
    \multicolumn{5}{@{}l}{\textit{Table~\ref{tab:master_verdict} (continued)}}\\
    \toprule
    \textbf{Item} & \textbf{Verdict} & \textbf{Status quo ante} & \textbf{Status 2026 / decisive evidence} & \textbf{What still blocks it} \\
    \midrule
    \endhead
    \midrule
    \multicolumn{5}{@{}r}{\textit{continued on next page}}\\
    \endfoot
    \bottomrule
    \endlastfoot
    Industrial parallel ablation &
    Collapsed &
    Single-spot serial ablation; throughput fixed by the per-spot rate &
    20-spot, 100~\W, 6~mm$^3$~min$^{-1}$ operation~\cite{lutzEfficientUltrashortPulsed2021} (spot-array geometry precedent: a 64-spot design study reporting no material, wavelength or power figures~\cite{hofmannDesignMultibeamOptics2020}); hardware demonstrated to 210~\W\ under pulsed loading~\cite{tangExtendingOperationalLimit2025}, with CW-only figures to 383~\W\ and 1.4~\kW\ not yet validated for this pulsed regime~\cite{zuoHighPerformanceNIRLaserBeam2025, wolenskiKilowattAveragePower2025} &
    Integrating 300-\W-class SLMs into the pipeline; present best remains 100~\W~\cite{lutzEfficientUltrashortPulsed2021, tangExtendingOperationalLimit2025} \\
    Parallel two-photon polymerisation &
    Collapsed (metalens rung single-study) &
    Single-focus serial writing at \um~s$^{-1}$; hours-to-days per part &
    Holographic and DOE-array routes each of order $10^{8}$~voxel~s$^{-1}$, independently reported by separate groups~\cite{zhangHighThroughputTwoPhoton3D2024, kieferMultiphoton77focus2024}; the $120{,}000$-focus metalens-array platform matches that rate ($1.2\times10^{8}$~voxel~s$^{-1}$) over a 12~cm$^2$ field but is a single group's result~\cite{gu3DNanolithographyMetalens2025} &
    Per-spot energy budget and long-duration operation at $> 100$ foci~\cite{zhangHighThroughputTwoPhoton3D2024}; independent replication of the metalens-array platform~\cite{gu3DNanolithographyMetalens2025} \\
    Sub-\textmu m photonic-device writing &
    Persists &
    Gaussian-spot writing; resolution limited by uncorrected spherical aberration at depth &
    Per-pass aberration penalty couples write speed to feature quality~\cite{jiaRecentProgressFemtosecond2023, heSlitBeamShaping2022, liuFabricationSinglemodeCircular2021} &
    In-line closed-loop aberration correction~\cite{weinackerIterativePreCompensation3D2023, wangIntegratedLCOSSLMBasedLaser2025} \\
    Long-duration high-power SLM reliability &
    Persists &
    Commodity air-cooled SLMs capped at tens of~\W &
    kW-class results are CW with customised cooling~\cite{wolenskiKilowattAveragePower2025, zuoHighPerformanceNIRLaserBeam2025} &
    A characterised 24/7 fs-burst duty cycle~\cite{tangExtendingOperationalLimit2025, buskeAdvancedBeamShaping2022} \\
    Peak-fluence tolerance of post-modulator static optics &
    Persists &
    Damage tolerance known only from ns-pulse LIDT testing &
    Metasurfaces and DOEs uncharacterised at fs-burst, MHz loads~\cite{wangVortexfieldEnhancementHighthreshold2024, crottiGiantUltrafastDichroism2024} &
    fs-burst damage-threshold characterisation at platform throughputs~\cite{gu3DNanolithographyMetalens2025, lafargueInVolumeGlassModification2024} \\
\end{longtable}
}

The seven-axis benchmark and the refresh-rate--average-power plane introduced here are offered as the instrument for tracking that shift as it reaches further regimes. Two of the review's findings share one root: the peak-fluence tolerance that bounds the collapse and the per-pattern computational cost of the seventh axis are both decisive yet rarely reported. We therefore close with a reporting standard rather than only a diagnosis. Every demonstration of programmable beam shaping for micromachining should report the sustained average power and fs-burst peak fluence at the modulator over a stated duty cycle, the optical efficiency and a beam-quality metric under that same load, and the per-pattern generation cost in wall-clock latency and energy, benchmarked for learned methods against the iterative Gerchberg--Saxton baseline. These populate the empty peak-fluence and compute columns of table~\ref{tab:six_axis_master}, turning the frame from a descriptive instrument into a comparative one.

\subsection{Milestones for the coming years}

Six concrete milestones follow, offered as predictions for the coming years rather than open aspirations.

\textbf{A kilowatt SLM at an industrial duty cycle.} The target is a cooled device that holds $>1$~\kHz\ refresh and $>95$~\% optical efficiency in sustained operation. The 1.4~\kW\ CW result already fixes the power, so what remains is a documented 24/7 cycle under femtosecond bursts rather than a continuous-wave peak \cite{wolenskiKilowattAveragePower2025, tangExtendingOperationalLimit2025}.

\textbf{Hologram generation in under a second for arbitrary 3D targets.} A single network that outputs the phase pattern for a $256 \times 256 \times 32$ voxel volume in under one second on a commodity GPU. It must do so without retraining for each new target \cite{yuUseDeepLearning2025, eybposhDeepCGH3DComputergenerated2020, shiRealtimePhotorealistic3D2021}.

\textbf{TPP beyond $10^{9}$ voxels per second.} A throughput above $10^{9}$~voxels~s$^{-1}$ at a voxel below 200~\nm. The obvious route is to combine the parallelism of a metalens array with the resolution of two-step absorption \cite{gu3DNanolithographyMetalens2025, messerShoeboxsized3DLaser2024}.

\textbf{Closed-loop aberration correction on the production line.} An automatic SLM correction, driven by a camera in the loop, that updates during processing rather than offline. SiC slicing and iterative precompensation already point the way \cite{wangIntegratedLCOSSLMBasedLaser2025, cuiLightPeopleProfessor2022, wangReviewFemtosecondLaser2024}.

\textbf{DOE design that builds in the damage threshold.} A machine-learning design loop whose loss function carries the substrate damage threshold as an explicit term. The optimiser then trades fidelity against survivability \cite{khoninaAdvancementsApplicationsDiffractive2024, liaoDifferentiableDesignFreeform2024}.

\textbf{A reconfigurable metasurface at industrial scale.} A device larger than 1~cm$^2$ with $>1$~\kHz\ refresh and a damage threshold compatible with femtosecond pulses. This means scaling up today's demonstrations, which run at square-millimetre size with picosecond pulses \cite{crottiGiantUltrafastDichroism2024, lengMetasurfaceMirrorsBased2024, chenLightdrivenPhaseTransition2024}.

\newpage
Two milestones offer the sharpest tests of the co-design thesis. Should a cooled kilowatt-class SLM degrade irreversibly within one production shift, the reading of reliability as an engineering-investment rather than a physical limit is falsified. Should a learned-CGH pipeline be benchmarked and found to consume more energy per pattern than the Gerchberg--Saxton baseline it replaces, the premise that co-design lowers rather than merely relocates the per-pattern cost is falsified.

\section{Conclusions and perspectives}
\label{sec:conclusions}

For two decades, programmable and static beam shaping have been treated as mutually exclusive regimes. That separation no longer holds in full. The two families were divided by an apparent trade-off between throughput and flexibility. Programmable modulators reconfigured freely but could not handle high average power, while static elements handled high power but were fixed once fabricated.

Seven hardware classes and five algorithm families were assessed across seven common axes, with each modulator read together with the algorithm that drives it. They converge on a single conclusion. Performance is governed by the co-design of the modulator, its algorithm and the process physics, not by the raw specification of any single component. On the average-power axis, that co-design has already dissolved the trade-off. In industrial parallel ablation and in high-throughput two-photon polymerisation, a programmable device now sustains, in continuous operation, an average power once reserved for static optics.

The result is bounded. It holds on the average-power axis alone, while the peak-fluence tolerance that ultimately governs micromachining remains uncharacterised across all programmable classes. Economics narrow it further. Static optics stay cheaper wherever the geometry is fixed, and the volume is high. A third limit is more fundamental. Once the pulse rate outruns the material's response time, beam shaping can no longer add throughput, because the bottleneck becomes the material rather than the optics.

Programmable shaping has therefore not displaced static optics. It has widened the range of conditions under which it is the better choice. The classes are better seen as complementary than competing. Each occupies a different region of the design space, and which one to use depends on the dominant constraint of the application. Where the frontier moved, it moved because a hardware advance and an algorithmic one arrived together.

The deeper contribution is conceptual. In past years, progress in beam shaping was measured by asking which optical technology performs better. That question is losing its force. Beam shaping is becoming an optical--computational discipline, in which performance is governed by system architecture rather than by any single component's specification. The bottleneck moves accordingly. Once a programmable device matches a static one in power and throughput, the binding constraint shifts from the beam shaper to the process it drives, whether the photochemistry of the resist, the thermomechanics of the substrate, or the energy cost of computing each pattern.


\section*{CRediT authorship contribution statement}
\textbf{K.~Kobliha:} Conceptualisation, Methodology, Investigation, Formal analysis, Writing (original draft), Visualisation. \textbf{P.~Hauschwitz:} Conceptualisation, Supervision, Writing (review and editing), Funding acquisition, Resources.

\section*{Declaration of competing interest}
The authors declare that they have no known competing financial interests or personal relationships that could have appeared to influence the work reported in this paper.

\section*{Funding}
This work was co-funded by the European Union and the state budget of the Czech Republic under the project LasApp CZ.02.01.01/00/22\_008/0004573.

\section*{Declaration of generative AI and AI-assisted technologies in the writing process}
During the preparation of this work, the authors used a large language model-based assistant (Claude, Anthropic) for language editing and drafting assistance. After using this tool, the authors reviewed and edited the content as needed and take full responsibility for the content of the publication.

\section*{Data availability}
No new data were created or analysed in this study. Data sharing is not applicable to this article.

\section*{Author biographies}
\noindent
\begin{minipage}[t]{0.16\textwidth}
\vspace{0pt}
\includegraphics[width=\linewidth]{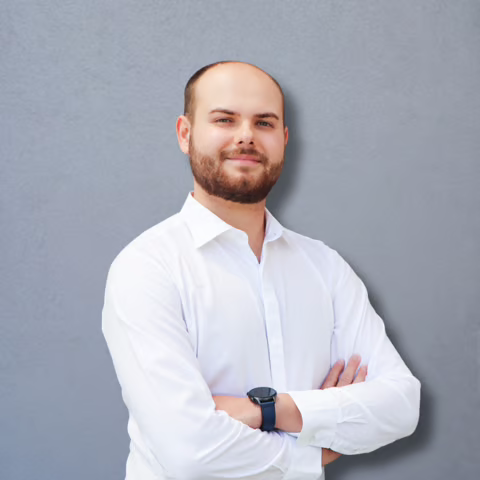}
\end{minipage}\hfill
\begin{minipage}[t]{0.80\textwidth}
\vspace{0pt}
\textbf{K.~Kobliha} is a PhD student at the HiLASE Centre (Institute of Physics, Czech Academy of Sciences) and at the Czech Technical University in Prague, where he completed his master's degree in 2026. He works on dynamic beam shaping for ultrafast laser micromachining, developing machine-learning and camera-in-the-loop methods to compute holograms, and on laser-textured functional surfaces.
\end{minipage}

\vspace{10pt}

\noindent
\begin{minipage}[t]{0.16\textwidth}
\vspace{0pt}
\includegraphics[width=\linewidth]{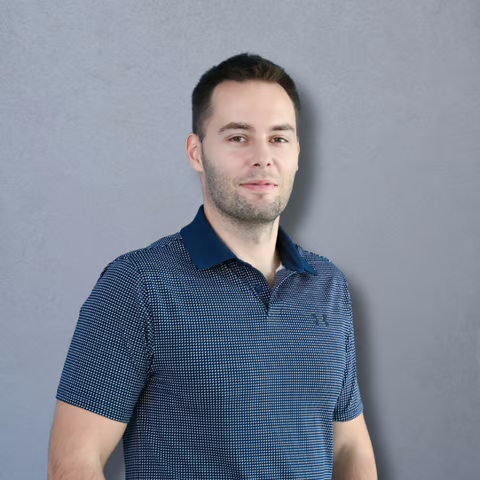}
\end{minipage}\hfill
\begin{minipage}[t]{0.80\textwidth}
\vspace{0pt}
\textbf{P.~Hauschwitz} is a group leader at the HiLASE Centre (Institute of Physics, Czech Academy of Sciences). His group develops high-throughput micro- and nanostructuring with ultrafast lasers, using multi-beam and beam-shaping methods such as diffractive optics and SLM-based parallel processing. A main theme is laser-induced periodic surface structures (LIPSS) that make surfaces anti-reflective, anti-adhesive, or biocompatible.
\end{minipage}

\newpage
\printbibliography

\end{document}